\documentclass[aps,pra,11pt]{revtex4-2}

\usepackage{amsmath,amssymb,bm,caption,setspace,xspace,fancyvrb,scrextend,siunitx,euscript,color,xcolor,ulem}
\usepackage{enumerate,braket,float,wrapfig,booktabs,textcomp,comment}
\usepackage[pdftex]{graphicx}
\usepackage[displaymath]{lineno}
\usepackage[colorlinks]{hyperref}
\usepackage{notes2bib}

\newcommand{\uu}[1]{\ensuremath{\, \mathrm{#1}}}
\newcommand{\chem}[1]{\ensuremath{\mathrm{#1}}}

\newcommand{\ra}{\ensuremath{\rightarrow}\xspace}
\let\v\bm

\def\sE{{\ensuremath{\mathcal E}}}

\graphicspath{{./Figures/}}

\begin{document}

\title{EuCl$_3\cdot 6$H$_2$O as the solid-state platform for searches for new P,T-odd physics}

\author{A.~O.~Sushkov}\email{asu@bu.edu}
\affiliation{Department of Physics, Boston University, Boston, MA 02215, USA}
\affiliation{Department of Electrical and Computer Engineering, Boston University, Boston, MA 02215, USA}
\affiliation{Photonics Center, Boston University, Boston, MA 02215, USA}

\begin{abstract}
Condensed matter systems offer the opportunity to maximize the number of spins, probed in magnetic resonance experiments that search for physics beyond the standard model. EuCl$_3\cdot 6$H$_2$O is a promising solid-state platform for such searches. The narrow inhomogeneous linewidth of the Eu optical $^7\hspace{-0.1em}F_0 \leftrightarrow\,^5\hspace{-0.1em}D_0$ transition in this material opens the opportunity to optically pump the $^{153}$Eu nuclear spin ensemble into a single hyperfine sublevel. The closely-spaced opposite-parity nuclear energy levels, and the low-energy collective octupole mode, or a possible static octupole deformation, enhance the $^{153}$Eu nuclear Schiff moment. The optimistic estimate of the effective electric field is $E^*=10\uu{MV/cm}$. With this estimate, the CASPEr-electric search for the gluon interaction of axion dark matter can reach QCD axion sensitivity with a 3~cm sample.
\end{abstract}

\maketitle

% Text spacing.
\baselineskip16pt

%\tableofcontents

\section{Introduction}

\noindent
A number of experimental searches for new fundamental physics are based on the magnetic resonance approach~\cite{Purcell1950,Safronova2018}. Such measurements search for energy level shifts or spin dynamics, induced by new particles or interactions. The number of spins probed by the experiment is one of the key parameters that determines sensitivity~\cite{Degen2017}.
Condensed matter systems offer the opportunity to work with macroscopic spin ensembles, on the order of Avogadro's number~\cite{Budker2006}.
Non-centrosymmetric solids are promising for searches for violations of parity (P) and time reversal (T) symmetries~\cite{Leggett1978,Mukhamedjanov2005a,Rushchanskii2010,Eckel2012}.
The primary motivation of the present work is the CASPEr-electric search for the electric dipole moment (EDM) and the gradient interactions of axion dark matter~\cite{Budker2014,Aybas2021a,Aybas2021b}.
This nuclear magnetic resonance (NMR)-based experiment searches for spin dynamics due to an oscillating spin energy shift.
However the approach described in this work can also be used to search for the static P,T-violating nuclear Schiff moment~\cite{Sushkov2023}.

The QCD axion is the best-motivated ultra-light dark matter candidate, thanks to its resolution of the strong-CP problem~\cite{Peccei1977,Weinberg1978,Wilczek1978,Kim2010a,Irastorza2018a}.
A number of experiments search for the electromagnetic and the gradient interactions of axion-like dark matter~\cite{Sikivie1983,Braine2020,Graham2018a,Ernst2018,Schutz2020,Graham2015a, Brubaker2017b, McAllister2017, Melcon2018, Alesini2019a, Lee2020, Sikivie2014b, Chaudhuri2015, Kahn2016, Chaudhuri2018, Ouellet2019,Garcon2019b,Wu2019a,Crescini2020,Sikivie2021,Gramolin2021,DMRadioCollaboration2023}.
However it is the interaction with the gluon field that is the defining model-independent coupling of the QCD axion, that solves the strong-CP problem~\cite{Graham2011}. Any claim of QCD axion detection would necessitate a search for this EDM interaction.
In this work we propose to use \chem{EuCl_3\cdot 6H_2O} as the non-centrosymmetric crystal for the CASPEr-e search, with the potential to reach QCD axion sensitivity.
This material contains no unpaired electron spins. We focus on the ensemble of $^{153}$Eu atoms (52\% natural abundance), with nuclear spin $I=5/2$.
The EDM interaction gives rise to an energy shift 
\begin{align}
	\label{calE1}
	\delta \sE= -d_n\v{E}^*\cdot\v{I}/I,
\end{align}
where $d_n$ is the neutron EDM and $E^*$ is the effective electric field, that determines the sensitivity to P,T-odd physics~\cite{Ludlow2013,Skripnikov2016,Flambaum2020,Aybas2021a}.
We note that $E^*$ is not a real electric field, in the sense that it is not sourced by electric charges and does not obey Maxwell's equations. It does have the same dimensions and the same discrete transformation properties as an electric field. In Ref.~\cite{Sushkov2023} we calculated the magnitude of the effective electric field in \chem{EuCl_3\cdot 6H_2O} to be 
\begin{align}
	E^*=10\uu{MV/cm}.
	\label{eq:energy50}
\end{align}
This is the optimistic estimate, based on the assumption that the $^{153}$Eu nuclear Schiff moment is enhanced by an octupole deformation~\cite{Dalton2023}. 

\begin{figure}[h!]
	\centering
	\includegraphics[width=8cm]{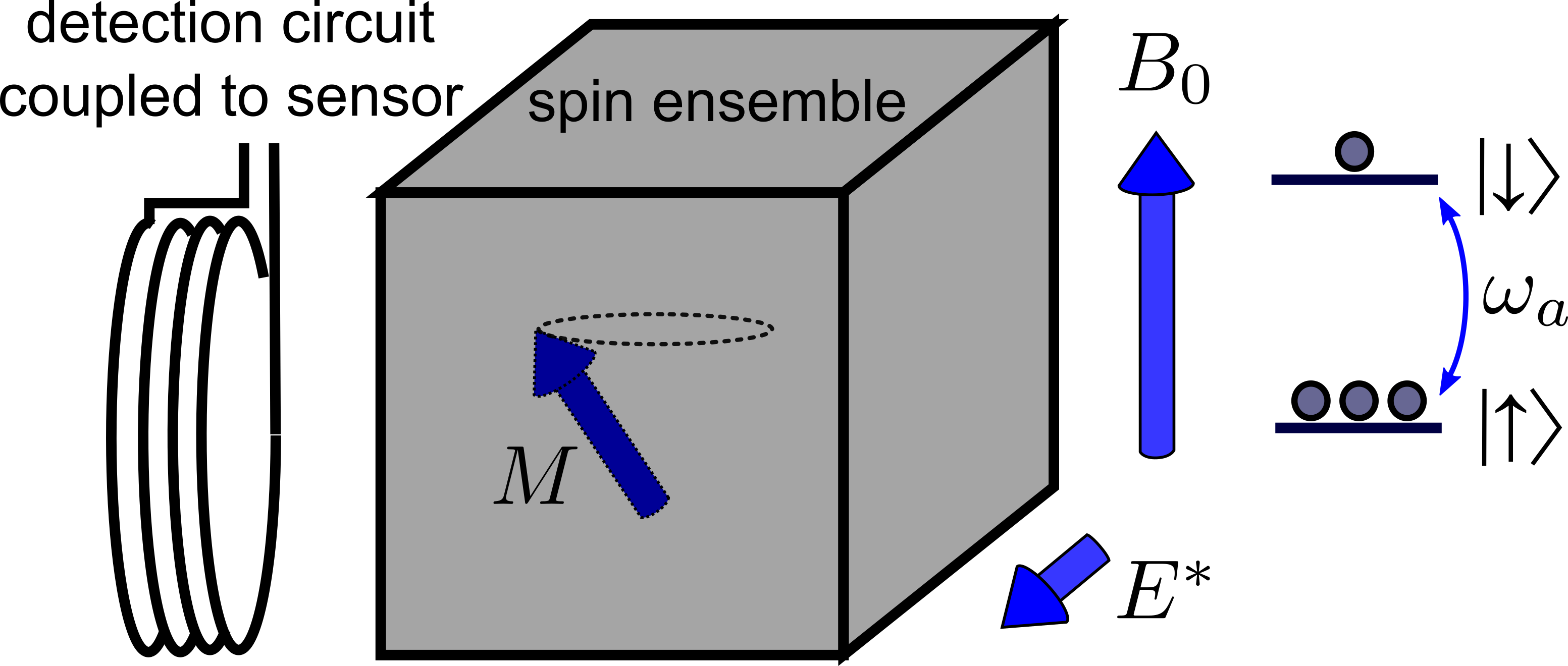}
	\caption{\fontsize{10}{11}\selectfont
		CASPEr-e scheme. 
	}
	\label{fig:CASPEr-e}
\end{figure}

The CASPEr-e experimental schematic is shown in Fig.~\ref{fig:CASPEr-e}.
The nuclear spins inside the sample are initialized with orientation perpendicular to $\v{E}^*$. The EDM interaction~\eqref{calE1} with axion dark matter creates an oscillating torque on the nuclear spins. We quantify the magnitude of this torque by the Rabi frequency $\Omega_a=d_nE^*/\hbar$. This torque tilts the spins, if it is resonant with their Larmor frequency. The experimental observable is the oscillating transverse magnetization:
\begin{align}
	M_a = uM_0\Omega_aT_2\cos{(\omega_a t)} = u(p\hbar\gamma n)(d_nE^*/\hbar)T_2\cos{(\omega_a t)},
	\label{eq:420}
\end{align}
where $M_0=p\hbar\gamma n$ is the magnetization of the spin ensemble with polarization fraction $p$ and number density $n$, $T_2$ is its spin coherence time, and $u$ is a dimensionless spectral factor that takes into account the inhomogeneous broadening of the spin ensemble and the detuning between the ALP Compton frequency and the spin Larmor frequency~\cite{Aybas2021a}.
Near resonance $uT_2\approx T_2^*$ is a good approximation.
In order to measure the transverse magnetization $M_a$, a pickup coil inductively couples the spin ensemble to a voltage or current sensor.
The material-dependent parameters that affect the sensitivity of the CASPEr-e search are $p$, $n$, $E^*$, and $T_2$. 
The first-generation CASPEr-e measurements focused on $^{207}$Pb$^{2+}$ ions, with nuclear spin $I=1/2$, in a poled ferroelectric PMN-PT crystal with the chemical formula: 
\chem{(PbMg_{1/3}Nb_{2/3}O_3)_{2/3}-(PbTiO_3)_{1/3}}~\cite{Aybas2021a}.
In this work we propose to use the $^{153}$Eu nuclear spin ensemble in \chem{EuCl_3\cdot 6H_2O} for the CASPEr-e search. We show that this material has the potential to make significant improvements in all these parameters, and estimate that it is possible to reach QCD axion sensitivity with a 3~cm sample.

\section{Properties of \chem{EuCl_3\cdot 6H_2O}}

Our experimental search exploits the remarkably narrow inhomogeneous linewidth of the $579.7\uu{nm}$ Eu $^7F_0\,\rightarrow\, ^5D_0$ optical transition
in \chem{EuCl_3\cdot 6H_2O}. Linewidths of $25\uu{MHz}$ has been observed in the stoichiometric crystal, isotopically purified in $^{35}$Cl~\cite{Ahlefeldt2016}. We plan to make use of this ultra-narrow optical linewidth to optically pump and hyperpolarize the $^{153}$Eu nuclear spin ensemble, improving $p$.
Due to proximity between opposite-parity nuclear levels of the $^{153}$Eu isotope and a potential collective quadrupole deformation, the nuclear Schiff moment is enhanced by a factor between $\approx5$ and $\approx100$, compared to $^{207}$Pb~\cite{Sushkov2023}. This leads to an improvement in $E^*$. The values of other relevant parameters are given in table~\ref{tab:1}.

%------------------------------------------
\begin{table}[h!]
	\centering
	\begin{tabular}{c|ll}
		& PMN-PT & \chem{EuCl_3\cdot6H_2O} \\ \midrule
		$E^*$ & $340\uu{kV/cm}$ & $10\uu{MV/cm}$ \\  \midrule
		$n$ & $3.4\times 10^{21}\uu{cm^{-3}}$ & $2\times 10^{21}\uu{cm^{-3}}$ \\  \midrule
		$p$ & $5\times 10^{-4}$ & $0.3$ (proj.) \\ \midrule
		$T_2$ & $(16.7\pm 0.9)\uu{ms}$ & $40\uu{ms}$~\cite{Ahlefeldt2013b} \\ \bottomrule
	\end{tabular}
	\caption{\fontsize{10}{11}\selectfont
		Nuclear spin ensemble parameters for PMN-PT~\cite{Aybas2021a} vs \chem{EuCl_3\cdot6H_2O} (this work).}\label{tab:1}
\end{table} 
%------------------------------------------
\begin{figure}[h!]
	\centering
	\includegraphics[width=0.8\textwidth]{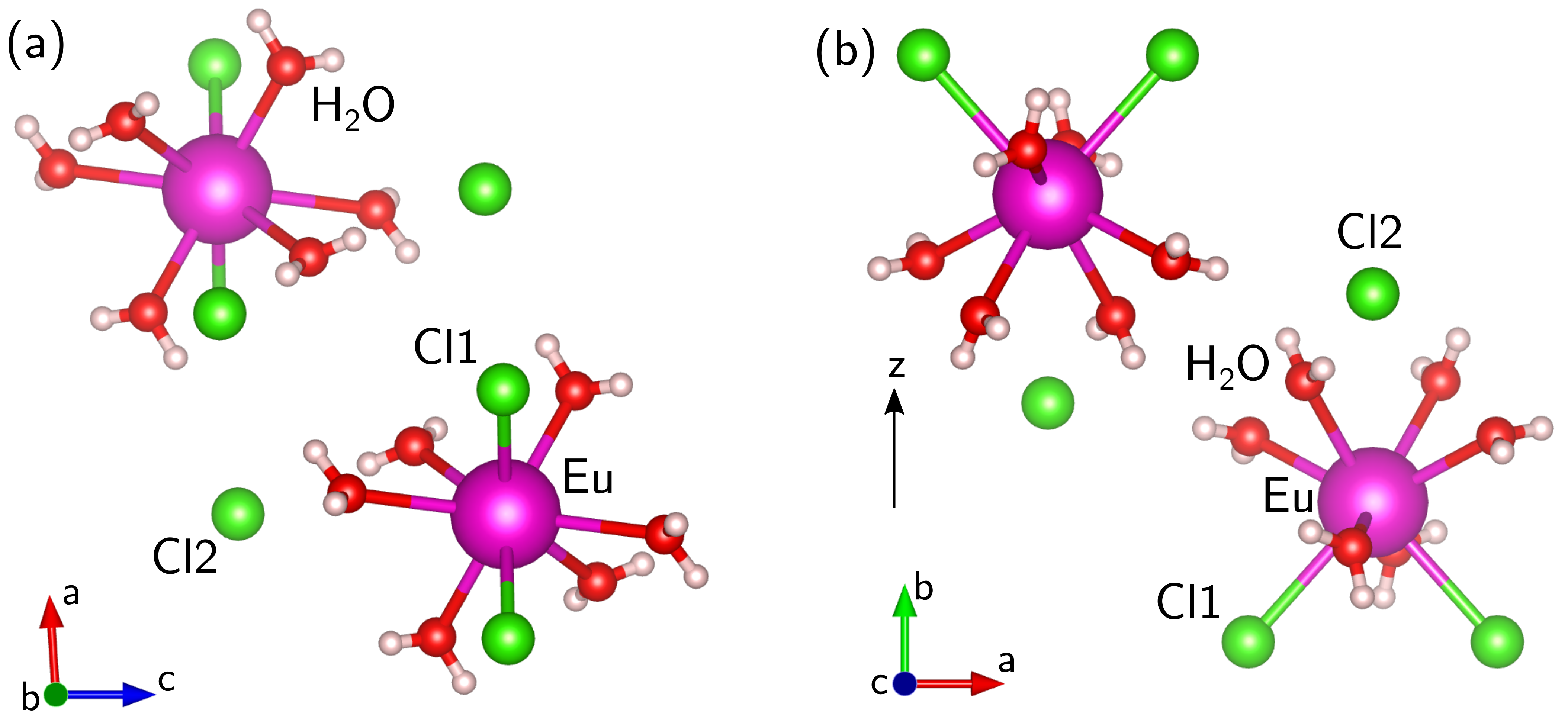}
	\caption{\fontsize{10}{11}\selectfont
		The structure of the unit cell of \chem{EuCl_3\cdot 6H_2O}~\cite{Tambornino2014}. (a) The projection viewed down the crystal b-axis. (b) The projection viewed down the crystal c-axis. Note that the Chlorine Cl2 atoms are significantly displaced along the c-axis with respect to the Eu atoms, thus they do not suppress the Eu site asymmetry.}
	\label{fig:200}
\end{figure}

\section{The $^{153}\rm{Eu}$ hyperfine structure in \chem{EuCl_3\cdot6H_2O}}

\begin{figure}[b!]
	%\vspace{-2mm}
	\centering
	\includegraphics[width=8cm]{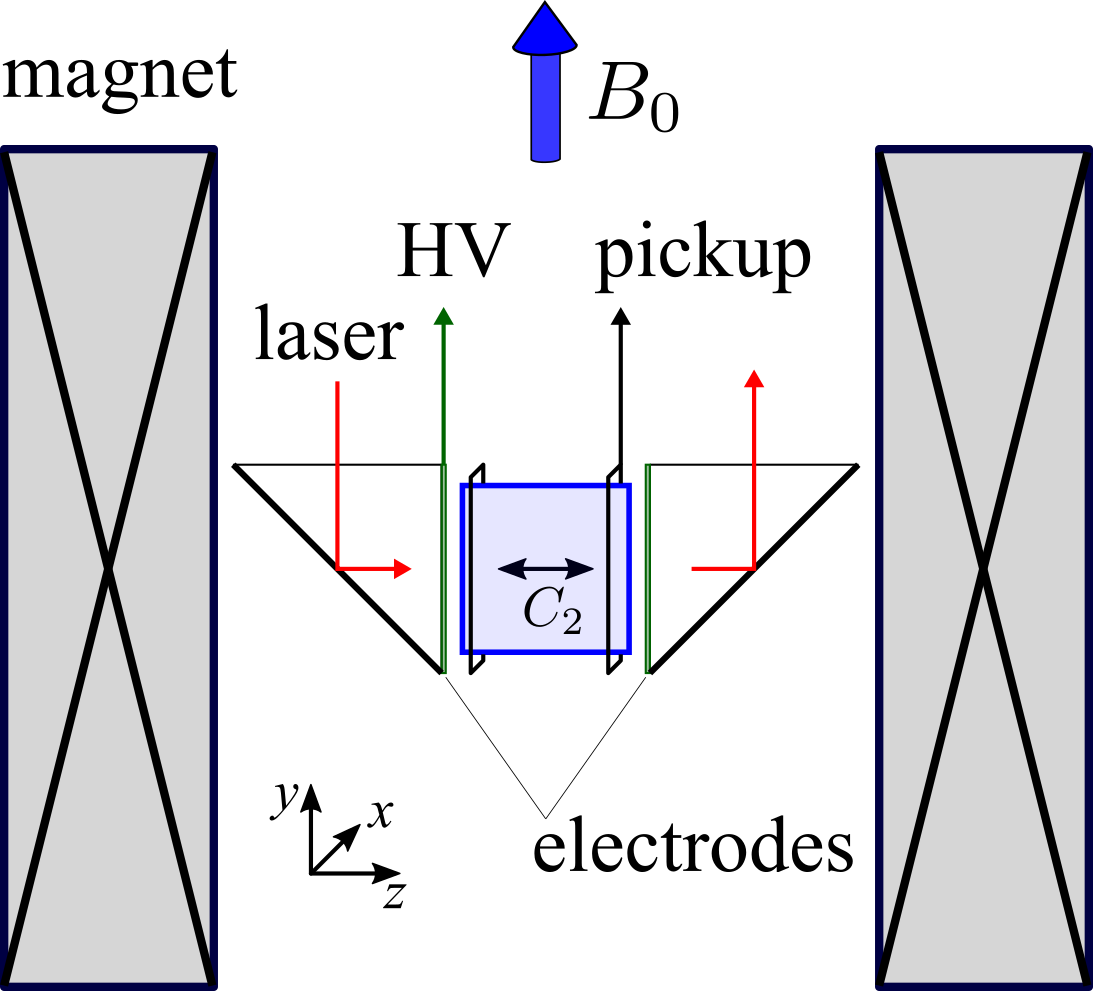}
	\caption{\fontsize{10}{11}\selectfont
		The experimental schematic for an axion search experiment with \chem{EuCl_3\cdot6H_2O}.
	}
	\label{fig:exp}
	%\vspace{-0.1cm}
\end{figure}

\noindent
In order to evaluate the feasibility of optical pumping and the CASPEr-e measurement scheme, we performed numerical simulations of the hyperfine structure, the Zeeman effect, and the optical spectra of the $^7\hspace{-0.1em}F_0\ra\hspace{0.1em}^5\hspace{-0.1em}D_0$ transition in \chem{^{153}EuCl_3\cdot6H_2O}. 
The experimental schematic and the coordinate axes are shown in Fig.~\ref{fig:exp}.

The effective nuclear spin Hamiltonian is given by
\begin{align}
	H = \bm{B}\cdot\bm{M}\cdot\bm{\hat{I}} + \bm{\hat{I}}\cdot\bm{Q}\cdot\bm{\hat{I}},
	\label{eq:ham1}
\end{align}
where $\bm{B}$ is the applied magnetic field, $\bm{M}$ is the nuclear Zeeman tensor, $\bm{\hat{I}}$ is the nuclear spin operator, and $\bm{Q}$ is the nuclear quadrupole tensor~\cite{Smith2022}. The orientation of the principal axes of these tensors relative to the laboratory coordinate system is given by the rotation matrix
\begin{align}
	R(\alpha,\beta,\gamma) = 
	\begin{bmatrix}
		\cos\alpha & -\sin\alpha & 0 \\
		\sin\alpha & \cos\alpha & 0 \\
		0 & 0 & 1
	\end{bmatrix}\cdot
	\begin{bmatrix}
		\cos\beta & 0 & \sin\beta \\
		0 & 1 & 0 \\
		-\sin\beta & 0 & \cos\beta
	\end{bmatrix}\cdot
	\begin{bmatrix}
		\cos\gamma & -\sin\gamma & 0 \\
		\sin\gamma & \cos\gamma & 0 \\
		0 & 0 & 1
	\end{bmatrix},
	\label{eq:R1}
\end{align}
defined in terms of the three Euler angles $\alpha,\beta,\gamma$. The C$_2$ symmetry of the crystal constrains one of the tensor principal axes to be along the C$_2$ axis, whose orientation relative to the laboratory coordinate system is described by the Euler angles $\alpha=\beta=0$, since we set the z-axis to be along the $C_2$ axis of the crystal, see fig.~\ref{fig:exp}. The tensors in the laboratory reference frame are given by:
\begin{align}
	\bm{M} = R(\alpha,\beta,\gamma_M)\cdot
	\begin{bmatrix}
		g_x & 0 & 0 \\
		0 & g_y & 0 \\
		0 & 0 & g_z
	\end{bmatrix}\cdot R^T(\alpha,\beta,\gamma_M),
	\label{eq:M1}
\end{align}
and
\begin{align}
	\bm{Q} = R(\alpha,\beta,\gamma_Q)\cdot
	\begin{bmatrix}
		-F & 0 & 0 \\
		0 & F & 0 \\
		0 & 0 & D
	\end{bmatrix}\cdot R^T(\alpha,\beta,\gamma_Q).
	\label{eq:M1}
\end{align}
Ref.~\cite{Smith2022} contains a detailed discussion of experimental measurements and calculations of these parameters, whose values are listed in table~\ref{tab:3}.
We do not include the quadratic Zeeman effect, since we will only consider hyperfine sublevel transitions.
\begin{table}[b!]
	\begin{center}
		\begin{tabular}{c|c|c}
			parameter & ground state  $^7\hspace{-0.1em}F_0$ & excited state $^5\hspace{-0.1em}D_0$ \\
			\hline 
			$\alpha$ ($\deg$) & $0$ & $0$ \\
			$\beta$ ($\deg$) & $0$ & $0$ \\
			$\gamma_M$ ($\deg$) & $22.69\pm0.07$ & $3.1\pm0.7$ \\
			$\gamma_Q$ ($\deg$) & $2.93\pm0.13$ & $13.22\pm0.03$ \\
			$g_x$ (MHz/T) & $\pm1.794\pm0.002$ & $+4.269\pm0.003$ \\
			$g_y$ (MHz/T) & $\mp0.673\pm0.002$ & $+4.105\pm0.003$ \\
			$g_z$ (MHz/T) & $+1.359\pm0.004$ & $+4.428\pm0.007$ \\
			$F$ (MHz) & $13.660932\pm0.000008$ & $23.21\pm0.02$ \\
			$D$ (MHz) & $0.791815\pm0.000014$ & $-4.80\pm0.02$ \\
			\hline 
		\end{tabular}
		\caption{Spin Hamiltonian parameters for the optical ground state  $^7\hspace{-0.1em}F_0$ and excited state $^5\hspace{-0.1em}D_0$ of $^{153}$Eu$^{3+}$ in \chem{EuCl_3\cdot6H_2O}~\cite{Smith2022}.
		}
		\label{tab:3}
	\end{center}
	\nonumber
\end{table} 

\begin{figure}[b!]
	%\vspace{-2mm}
	\includegraphics[width=\textwidth]{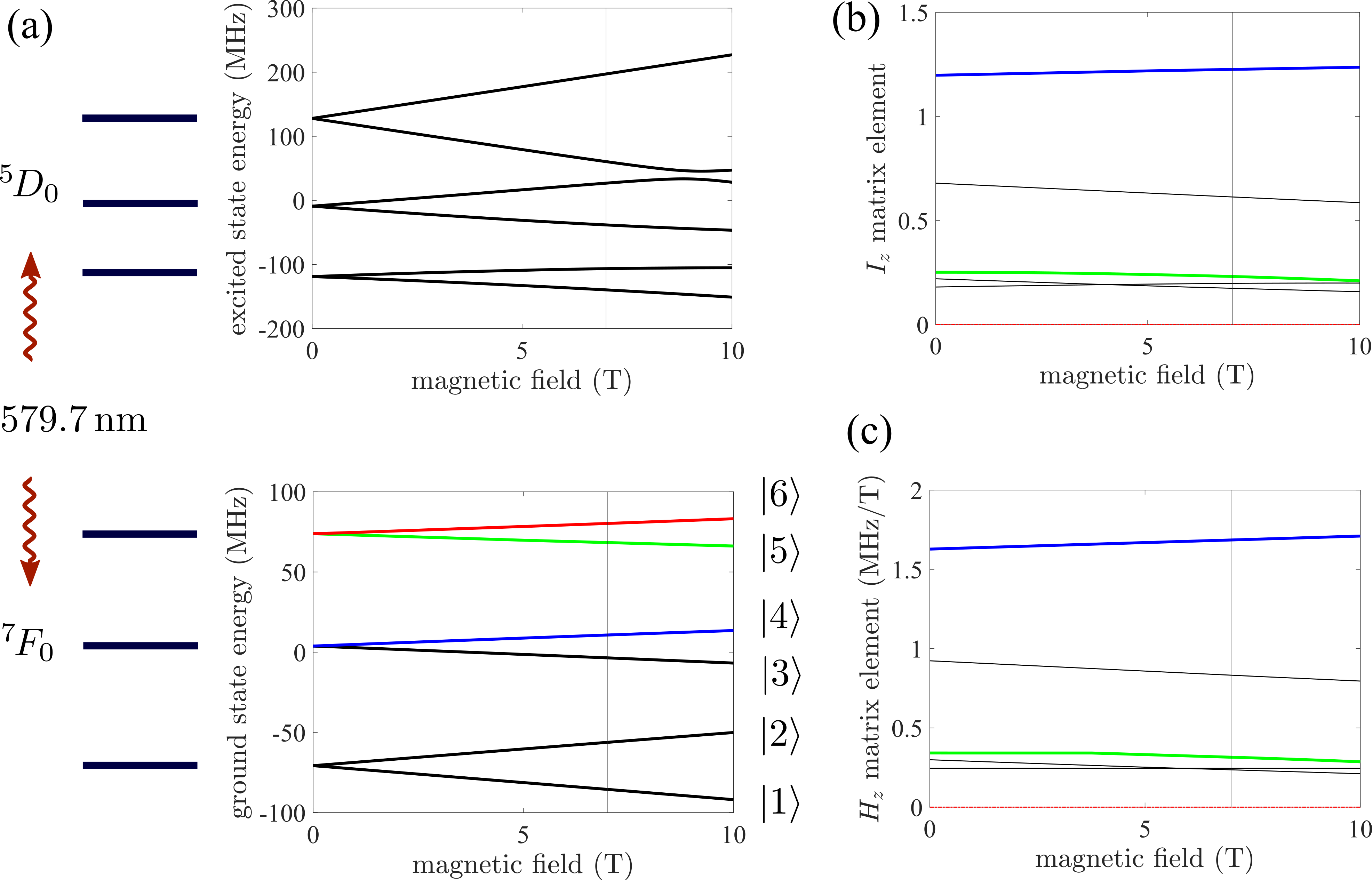}
	\caption{\fontsize{10}{11}\selectfont
		The hyperfine structure and nuclear spin Zeeman effect in $^{153}$Eu $^7\hspace{-0.1em}F_0$ and $^5\hspace{-0.1em}D_0$ states in \chem{EuCl_3\cdot6H_2O}. (a) The optical transition is at $579.7\uu{nm}$. The strong quadrupole interaction splits nuclear spin sublevels, which further split under magnetic field, applied perpendicular to the $C_2$ axis. Our objective is to achieve optical pumping into the ground state sublevel marked $|6\rangle$ and make use of the $|6\rangle\leftrightarrow|5\rangle$ transition. (b,c) Magnetic dipole matrix elements for ground-state nuclear spin transitions.
	}
	\label{fig:EuCl3levels}
	%\vspace{-0.1cm}
\end{figure}
For initial estimates, we set the applied magnetic field $\bm{B}_0$ to point perpendicular to the C$_2$ axis of the crystal, fig.~\ref{fig:exp}. 
%Note that laser light has to propagate along the C$_2$ axis, in order to minimize light absorption in the optically-dense sample.
We adopt the laboratory frame with the z-axis along the $C_2$ of the crystal and the y-axis along the magnetic field $\bm{B}_0$ -- this is the axis of the solenoid magnet.
We numerically diagonalize the Hamiltonian in eq.~\eqref{eq:ham1}, using the values of the parameters given in Tab.~\ref{tab:3}. The resulting ground state and excited state Zeeman energy levels are shown in fig.~\ref{fig:EuCl3levels}(a). Note that the nuclear spin projection is not a good quantum number, since the quadrupole and Zeeman interactions are of comparable strength.
We label the $^7\hspace{-0.1em}F_0$ hyperfine sublevels   $|1\rangle\rightarrow|6\rangle$.
%We plan to optically pump the ground-state population into the sublevel marked $|6\rangle$.

In the CASPer experimental approach the EDM interaction~\eqref{calE1} of the axion dark matter can drive transitions between ground-state nuclear spin sublevels, if their splitting matches the axion Compton frequency. 
We plan to optically pump the $^7\hspace{-0.1em}F_0$ ground-state population into one of the hyperfine sublevels, such as $|6\rangle$.
The search for the unknown axion Compton frequency is performed by tuning the Zeeman sublevel splitting with the applied magnetic field $B_0$.
%After the optical pumping is complete, a resonant $\pi$-pulse can transfer ground-state population into any hyperfine sublevel.
The magnetic dipole transition matrix elements between the hyperfine sublevels are important parameters for sensitivity and calibration of our axion dark matter search.
Since the effective electric field is in the z-direction, the detected signal is proportional to the product of matrix elements $\langle b|\hat{\bm{z}}\cdot\bm{M}\cdot\bm{\hat{I}}|a\rangle\langle b|\hat{I}_z|a\rangle$, where $|a\rangle$ and $|b\rangle$ are the two hyperfine spin states under study.
We calculate matrix elements of this Hamiltonian, using the eigenvectors of the diagonalized Hamiltonian~\eqref{eq:ham1}. Matrix elements for transitions starting in state $|6\rangle$ are shown in fig.~\ref{fig:EuCl3levels}(b,c). We note that some of the matrix elements are suppressed, but there is always a pair of sublevels where there is no suppression.
We verified that in the large magnetic field limit, when the Zeeman interaction dominates over the quadrupole interaction, the matrix elements correspond to the usual magnetic dipole-allowed transitions.

The results shown in Fig.~\ref{fig:EuCl3levels} are for the specific crystal alignment shown in Fig.~\ref{fig:exp}. The energy level structure and the transition matrix elements can be tuned by rotating the crystal with respect to the applied field $B_0$. This degree of freedom can be used to optimize the experimental sensitivity for a particular search frequency range. In this way, with $B_0$ limited to 7~T, the search for axion dark matter can cover frequencies up to 160~MHz.

\section{Nuclear spin polarization by optical pumping}

\noindent
Let us the discuss the scheme for optically pumping the nuclear spin ensemble into the $|6\rangle$ ground state sublevel. We rely on two properties of the \chem{EuCl_3\cdot6H_2O} crystal. The first is the narrow optical inhomogeneous linewidth, which has been demonstrated to be $25\uu{MHz}$ in a crystal isotopically purified with \chem{^{35}Cl}~\cite{Ahlefeldt2016}. This enables optical addressing of order-unity fraction of \chem{^{153}Eu} atoms. The second property is the hours-long nuclear spin population lifetime, observed in this material at cryogenic temperature~\cite{Ahlefeldt2016}. This enables long measurement timescales, after the optical pumping step. In addition, this eases the requirement on the pump laser power: assuming one scattered photon per atom, we can pump $10^{22}$ Eu atoms in one hour with $100\uu{mW}$ light power. The dominant non-radiative decay $^5\hspace{-0.1em}D_0\ra\hspace{0.1em}^7\hspace{-0.1em}F_0$ will lead to significant sample heating, which will make good thermal anchoring a key technical requirement. Importantly, the sample can be held at 4~K.

\begin{figure}[t!]
	\centering
	\includegraphics[width=\textwidth]{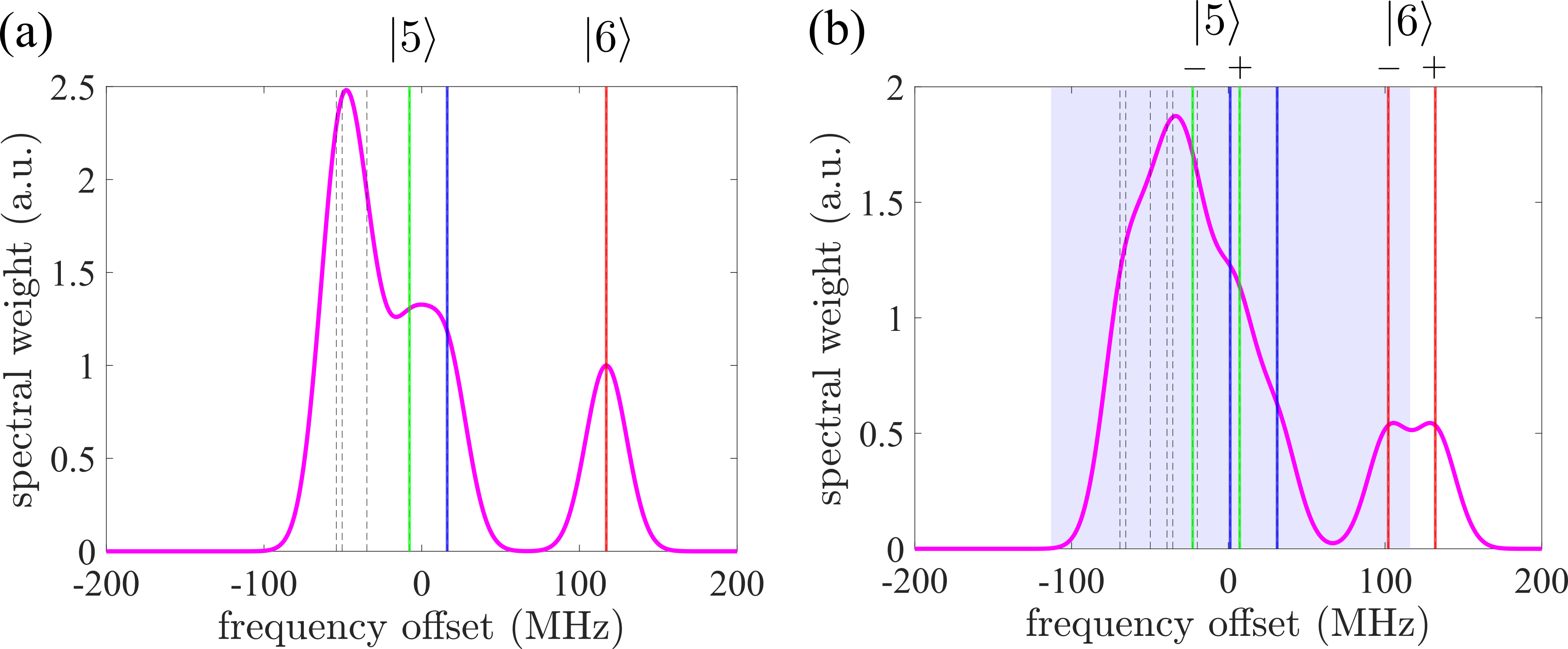}
	\caption{\fontsize{10}{11}\selectfont
		Simulated inhomogeneous optical spectrum of the $^7\hspace{-0.1em}F_0 \leftrightarrow\, ^5\hspace{-0.1em}D_0$ transition in \chem{^{153}EuCl_3\cdot6H_2O} at $B=7\uu{T}$. We assume diagonal nuclear sublevel transitions and $13\uu{MHz}$ inhomogeneous broadening. (a) Spectrum at no applied electric field. The ensemble can be pumped into the $|6\rangle$ sublevel by engineering the excitation spectrum to span $-100\uu{MHz}$ to $50\uu{MHz}$ (b) Spectrum with $10\uu{kV/cm}$ applied electric field. The transition frequencies of the two Eu non-centrosymmetric sites are marked with a `$+$' or a `$-$'. To pump the population into the $|6\rangle^+$ sublevel, we engineer the excitation spectrum to span $-100\uu{MHz}$ to $120\uu{MHz}$.
	}
	\label{fig:EuCl3spectrum}
	%\vspace{-0.1cm}
\end{figure}

We simulated the inhomogeneously-broadened optical spectrum of \chem{^{153}EuCl_3\cdot6H_2O} at $B_0=7\uu{T}$ by convolving each ground-to-excited state transition frequency with a Gaussian lineshape of width $13\uu{MHz}$, fig.~\ref{fig:EuCl3spectrum}(a).
In order to simplify the spectrum, we show only diagonal sublevel transitions; off-diagonal terms appear to contribute $\approx 20\%$ of the spectral weight~\cite{Ahlefeldt2016,Smith2022}. Optical pumping into the $|6\rangle$ ground state sublevel could proceed by engineering the laser excitation spectrum so that it does not excite $|6\rangle$ but covers the other transitions.

An important factor is the spatial asymmetry of the Eu ion sites in the crystal. Each Eu ion occupies a non-centrosymmetric site, see fig.~\ref{fig:200}. This is what gives rise to the effective electric field, calculated in Ref.~\cite{Sushkov2023}. However the two Eu sites in the unit cell are symmetric conjugates of each other, which means that they experience equal and opposite effective electric fields. If we addressed the entire \chem{^{153}Eu} spin ensemble, the net effect on the ensemble spin dynamics would vanish. 

Our approach is to selectively hyperpolarize the nuclear spins of one half of the \chem{^{153}Eu} spin ensemble - only those Eu ions that occupy one of the two non-centrosymmetric sites. In order to do this, we will apply an electric field $E_0$ along the C$_2$ axis of the crystal. This electric field will shift in opposite directions the optical transitions of the Eu ions in the two sites. Quantitatively, the optical transition frequency shift is given by
\begin{align}
	\Delta f = -\bm{D}\cdot \bm{E}_0,
	\label{eq:f1}
\end{align}
where $\bm{D}$ is proportional to the difference between the ground and excited state electric dipole moments, and includes the local field correction due to the crystal dielectric constant. We neglect the quadratic Stark shift. Measurements in ref.~\cite{Ahlefeldt2013c} result in the value $D = 1.5\uu{kHz/(V/cm)} \approx 10^{-3}\,ea_0$. 

If we apply $E_0 = 10\uu{kV/cm}$, we will shift the optical transitions of each of the two Eu sites by $\Delta f=\pm 15\uu{MHz}$. The simulated optical spectrum with the applied electric field is shown in fig.~\ref{fig:EuCl3spectrum}(b). The overall spectral envelope appears to be broadened. Vertical lines show the shifted transition frequencies of each nuclear spin sublevel, and each Eu ion site, with spatial symmetry marked with a `$+$' or a `$-$'. To pump the population into the $|6\rangle^+$ sublevel, we apply the electric field and excite the ensemble with the laser frequency spectrum spanning $(-100,120)\uu{MHz}$ frequency range (light blue shaded region in fig.~\ref{fig:EuCl3spectrum}(b)). The result is optical pumping into the $|6\rangle$ spin sublevel of only one of the Eu sites.

\section{Experimental sensitivity and systematics}

\noindent
The evaluation of experimental sensitivity follows the steps in Refs.~\cite{Aybas2021a,Aybas2021b,Sushkov2023}. Let us estimate the experimental signal created by the QCD axion dark matter, corresponding to the neutron EDM amplitude $d_n \approx 10^{-34}\uu{e\cdot cm}$, and the QCD $\theta$-parameter oscillation amplitude $\theta\approx 4\times10^{-19}$~\cite{Graham2013}. We assume the optimistic estimate of $E^*=10\uu{MV/cm}$ for $^{153}$Eu. The nuclear spin Rabi frequency is
\begin{align}
	\Omega_a = d_nE^*/\hbar\approx 2\times10^{-12}\uu{s^{-1}}.
	\label{eq:e01}
\end{align}
The corresponding steady-state tilt angle of the ensemble spin vector is $\Omega_aT_2\approx 10^{-13}$. 
Let us assume that the optical pumping approach achieves $p=0.3$ polarization of the nuclear spin ensemble. Then the resulting transverse magnetization is 
\begin{align}
	\mu_0M_a\approx 10^{-19}\uu{T}.
	\label{eq:e02}
\end{align}
One possibility is to detect the spin tilt via optical spectroscopy. However this method is limited by photon shot noise of the probe beam. Therefore let us consider the inductive detection method, as in Refs.~\cite{Budker2014,Aybas2021a}. We use a resonant pickup circuit, coupled to a SQUID current sensor. We assume the circuit quality factor $Q\approx3\times10^4$ and noise temperature $T_n\approx100\uu{mK}$. At these parameters the circuit Johnson-Nyquist noise is of the same order as the ensemble spin projection noise~\cite{Aybas2021b}. For the sample radius of $r_s=3\uu{cm}$ we estimate that we can achieve the sensitivity at the level given by Eq.~\eqref{eq:e02} after 3 minutes of averaging. In a scanning search for QCD axion dark matter this corresponds to covering a frequency octave in approximately two months.

The CASPEr-e approach to searching for axion dark matter has several natural ways to identify and reject systematic effects. Our initial experiments have shown that the main systematics are due to radiofrequency interference, which can be rejected by requiring that the candidate signals only appear when they are within the NMR linewidth of the spin Larmor frequency~\cite{Aybas2021a}. An additional possibility is to design the experiment with two detection channels, with opposite $E^*$ orientations, so that the axion dark matter signals appear out of phase~\cite{Gramolin2021}. Finally, the $^{151}$Eu isotope (48\% natural abundance) can be used as the co-magnetometer, since its nuclear Schiff moment is estimated to be a factor of 3 smaller than for $^{153}$Eu~\cite{Sushkov2023}.
This would, however, require a more complex optical pumping and detection scheme.

The hyperpolarized $^{153}$Eu spin ensemble in \chem{EuCl_3\cdot 6H_2O} can also be used to search for static P,T-violating effects, such as the nuclear Schiff moment~\cite{Budker2006,Sushkov2023}. The calculation of the statistical sensitivity is identical to the estimate given above. However, such solid state-based experiments need to incorporate much more careful studies of systematics~\cite{Sushkov2010,Rushchanskii2010,Eckel2012}. The systematic rejection methods listed above make \chem{EuCl_3\cdot 6H_2O} a promising platform.

\section{Conclusion}

\noindent
This work outlines the experimental approach to searching for the EDM interaction of axion dark matter, using the crystal \chem{EuCl_3\cdot 6H_2O}.
The narrow inhomogeneous linewidth of the optical $^7\hspace{-0.1em}F_0 \leftrightarrow\, ^5\hspace{-0.1em}D_0$ transition in this material enables optical pumping of the $^{153}$Eu nuclear spin ensemble. 
The closely-spaced opposite-parity nuclear energy levels, and the low-energy collective octupole mode 3$^-$, or even a static octupole deformation, enhance the $^{153}$Eu nuclear Schiff moment. The optimistic estimate of the effective electric field is $E^*=10\uu{MV/cm}$~\cite{Sushkov2023}. These features make \chem{EuCl_3\cdot 6H_2O} a promising platform to search for interactions violate the discrete parity and time-reversal symmetries. Incorporating this platform into the CASPEr-e search for the EDM interaction of axion dark matter enables reaching sensitivity to QCD axion dark matter with a 3~cm sample.
Ongoing experimental work is exploring this approach.

\begin{acknowledgments}

\noindent
I thank O. P. Sushkov, A. Yaresko, and R. Ahlefeldt for valuable discussions.
I acknowledge support by the National Science Foundation CAREER grant PHY-2145162, and the U.S. Department of Energy, Office of High Energy Physics program under the QuantISED program, FWP 100667.

\end{acknowledgments}

\bibliographystyle{apsrev4-2}
%\bibliography{library} 

\begin{thebibliography}{51}%
	\makeatletter
	\providecommand \@ifxundefined [1]{%
		\@ifx{#1\undefined}
	}%
	\providecommand \@ifnum [1]{%
		\ifnum #1\expandafter \@firstoftwo
		\else \expandafter \@secondoftwo
		\fi
	}%
	\providecommand \@ifx [1]{%
		\ifx #1\expandafter \@firstoftwo
		\else \expandafter \@secondoftwo
		\fi
	}%
	\providecommand \natexlab [1]{#1}%
	\providecommand \enquote  [1]{``#1''}%
	\providecommand \bibnamefont  [1]{#1}%
	\providecommand \bibfnamefont [1]{#1}%
	\providecommand \citenamefont [1]{#1}%
	\providecommand \href@noop [0]{\@secondoftwo}%
	\providecommand \href [0]{\begingroup \@sanitize@url \@href}%
	\providecommand \@href[1]{\@@startlink{#1}\@@href}%
	\providecommand \@@href[1]{\endgroup#1\@@endlink}%
	\providecommand \@sanitize@url [0]{\catcode `\\12\catcode `\$12\catcode
		`\&12\catcode `\#12\catcode `\^12\catcode `\_12\catcode `\%12\relax}%
	\providecommand \@@startlink[1]{}%
	\providecommand \@@endlink[0]{}%
	\providecommand \url  [0]{\begingroup\@sanitize@url \@url }%
	\providecommand \@url [1]{\endgroup\@href {#1}{\urlprefix }}%
	\providecommand \urlprefix  [0]{URL }%
	\providecommand \Eprint [0]{\href }%
	\providecommand \doibase [0]{https://doi.org/}%
	\providecommand \selectlanguage [0]{\@gobble}%
	\providecommand \bibinfo  [0]{\@secondoftwo}%
	\providecommand \bibfield  [0]{\@secondoftwo}%
	\providecommand \translation [1]{[#1]}%
	\providecommand \BibitemOpen [0]{}%
	\providecommand \bibitemStop [0]{}%
	\providecommand \bibitemNoStop [0]{.\EOS\space}%
	\providecommand \EOS [0]{\spacefactor3000\relax}%
	\providecommand \BibitemShut  [1]{\csname bibitem#1\endcsname}%
	\let\auto@bib@innerbib\@empty
	%</preamble>
	\bibitem [{\citenamefont {Purcell}\ and\ \citenamefont
		{Ramsey}(1950)}]{Purcell1950}%
	\BibitemOpen
	\bibfield  {author} {\bibinfo {author} {\bibfnamefont {E.~M.}\ \bibnamefont
			{Purcell}}\ and\ \bibinfo {author} {\bibfnamefont {N.~F.}\ \bibnamefont
			{Ramsey}},\ }\href {https://doi.org/10.1103/PhysRev.78.807} {\bibfield
		{journal} {\bibinfo  {journal} {Physical Review}\ }\textbf {\bibinfo {volume}
			{78}},\ \bibinfo {pages} {807} (\bibinfo {year} {1950})}\BibitemShut
	{NoStop}%
	\bibitem [{\citenamefont {Safronova}\ \emph {et~al.}(2018)\citenamefont
		{Safronova}, \citenamefont {Budker}, \citenamefont {Demille}, \citenamefont
		{Kimball}, \citenamefont {Derevianko},\ and\ \citenamefont
		{Clark}}]{Safronova2018}%
	\BibitemOpen
	\bibfield  {author} {\bibinfo {author} {\bibfnamefont {M.~S.}\ \bibnamefont
			{Safronova}}, \bibinfo {author} {\bibfnamefont {D.}~\bibnamefont {Budker}},
		\bibinfo {author} {\bibfnamefont {D.}~\bibnamefont {Demille}}, \bibinfo
		{author} {\bibfnamefont {D.~F.}\ \bibnamefont {Kimball}}, \bibinfo {author}
		{\bibfnamefont {A.}~\bibnamefont {Derevianko}},\ and\ \bibinfo {author}
		{\bibfnamefont {C.~W.}\ \bibnamefont {Clark}},\ }\href
	{https://doi.org/10.1103/RevModPhys.90.025008} {\bibfield  {journal}
		{\bibinfo  {journal} {Reviews of Modern Physics}\ }\textbf {\bibinfo {volume}
			{90}},\ \bibinfo {pages} {025008} (\bibinfo {year} {2018})},\ \Eprint
	{https://arxiv.org/abs/1710.01833} {arXiv:1710.01833} \BibitemShut {NoStop}%
	\bibitem [{\citenamefont {Degen}\ \emph {et~al.}(2017)\citenamefont {Degen},
		\citenamefont {Reinhard},\ and\ \citenamefont {Cappellaro}}]{Degen2017}%
	\BibitemOpen
	\bibfield  {author} {\bibinfo {author} {\bibfnamefont {C.~L.}\ \bibnamefont
			{Degen}}, \bibinfo {author} {\bibfnamefont {F.}~\bibnamefont {Reinhard}},\
		and\ \bibinfo {author} {\bibfnamefont {P.}~\bibnamefont {Cappellaro}},\
	}\href {https://doi.org/10.1103/RevModPhys.89.035002} {\bibfield  {journal}
		{\bibinfo  {journal} {Reviews of Modern Physics}\ }\textbf {\bibinfo {volume}
			{89}},\ \bibinfo {pages} {035002} (\bibinfo {year} {2017})},\ \Eprint
	{https://arxiv.org/abs/1611.02427} {arXiv:1611.02427} \BibitemShut {NoStop}%
	\bibitem [{\citenamefont {Budker}\ \emph {et~al.}(2006)\citenamefont {Budker},
		\citenamefont {Lamoreaux}, \citenamefont {Sushkov},\ and\ \citenamefont
		{Sushkov}}]{Budker2006}%
	\BibitemOpen
	\bibfield  {author} {\bibinfo {author} {\bibfnamefont {D.}~\bibnamefont
			{Budker}}, \bibinfo {author} {\bibfnamefont {S.~K.}\ \bibnamefont
			{Lamoreaux}}, \bibinfo {author} {\bibfnamefont {A.~O.}\ \bibnamefont
			{Sushkov}},\ and\ \bibinfo {author} {\bibfnamefont {O.~P.}\ \bibnamefont
			{Sushkov}},\ }\href {https://doi.org/10.1103/PhysRevA.73.022107} {\bibfield
		{journal} {\bibinfo  {journal} {Physical Review A}\ }\textbf {\bibinfo
			{volume} {73}},\ \bibinfo {pages} {022107} (\bibinfo {year}
		{2006})}\BibitemShut {NoStop}%
	\bibitem [{\citenamefont {Leggett}(1978)}]{Leggett1978}%
	\BibitemOpen
	\bibfield  {author} {\bibinfo {author} {\bibfnamefont {A.~J.}\ \bibnamefont
			{Leggett}},\ }\href {https://doi.org/10.1103/PhysRevLett.41.586} {\bibfield
		{journal} {\bibinfo  {journal} {Physical Review Letters}\ }\textbf {\bibinfo
			{volume} {41}},\ \bibinfo {pages} {586} (\bibinfo {year} {1978})}\BibitemShut
	{NoStop}%
	\bibitem [{\citenamefont {Mukhamedjanov}\ and\ \citenamefont
		{Sushkov}(2005)}]{Mukhamedjanov2005a}%
	\BibitemOpen
	\bibfield  {author} {\bibinfo {author} {\bibfnamefont {T.~N.}\ \bibnamefont
			{Mukhamedjanov}}\ and\ \bibinfo {author} {\bibfnamefont {O.~P.}\ \bibnamefont
			{Sushkov}},\ }\href {https://doi.org/10.1103/PhysRevA.72.034501} {\bibfield
		{journal} {\bibinfo  {journal} {Physical Review A}\ }\textbf {\bibinfo
			{volume} {72}},\ \bibinfo {pages} {34501} (\bibinfo {year} {2005})},\ \Eprint
	{https://arxiv.org/abs/0411226} {arXiv:0411226 [physics]} \BibitemShut
	{NoStop}%
	\bibitem [{\citenamefont {Rushchanskii}\ \emph {et~al.}(2010)\citenamefont
		{Rushchanskii}, \citenamefont {Kamba}, \citenamefont {Goian}, \citenamefont
		{Vanek}, \citenamefont {Savinov}, \citenamefont {Prokleska}, \citenamefont
		{Nuzhnyy}, \citenamefont {Knzek}, \citenamefont {Laufek}, \citenamefont
		{Eckel}, \citenamefont {Lamoreaux}, \citenamefont {Sushkov}, \citenamefont
		{Lezaic},\ and\ \citenamefont {Spaldin}}]{Rushchanskii2010}%
	\BibitemOpen
	\bibfield  {author} {\bibinfo {author} {\bibfnamefont {K.~Z.}\ \bibnamefont
			{Rushchanskii}}, \bibinfo {author} {\bibfnamefont {S.}~\bibnamefont {Kamba}},
		\bibinfo {author} {\bibfnamefont {V.}~\bibnamefont {Goian}}, \bibinfo
		{author} {\bibfnamefont {P.}~\bibnamefont {Vanek}}, \bibinfo {author}
		{\bibfnamefont {M.}~\bibnamefont {Savinov}}, \bibinfo {author} {\bibfnamefont
			{J.}~\bibnamefont {Prokleska}}, \bibinfo {author} {\bibfnamefont
			{D.}~\bibnamefont {Nuzhnyy}}, \bibinfo {author} {\bibfnamefont
			{K.}~\bibnamefont {Knzek}}, \bibinfo {author} {\bibfnamefont
			{F.}~\bibnamefont {Laufek}}, \bibinfo {author} {\bibfnamefont
			{S.}~\bibnamefont {Eckel}}, \bibinfo {author} {\bibfnamefont {S.~K.}\
			\bibnamefont {Lamoreaux}}, \bibinfo {author} {\bibfnamefont {A.~O.}\
			\bibnamefont {Sushkov}}, \bibinfo {author} {\bibfnamefont {M.}~\bibnamefont
			{Lezaic}},\ and\ \bibinfo {author} {\bibfnamefont {N.~A.}\ \bibnamefont
			{Spaldin}},\ }\href {https://doi.org/10.1038/nmat2799} {\bibfield  {journal}
		{\bibinfo  {journal} {Nature Materials}\ }\textbf {\bibinfo {volume} {9}},\
		\bibinfo {pages} {649} (\bibinfo {year} {2010})}\BibitemShut {NoStop}%
	\bibitem [{\citenamefont {Eckel}\ \emph {et~al.}(2012)\citenamefont {Eckel},
		\citenamefont {Sushkov},\ and\ \citenamefont {Lamoreaux}}]{Eckel2012}%
	\BibitemOpen
	\bibfield  {author} {\bibinfo {author} {\bibfnamefont {S.}~\bibnamefont
			{Eckel}}, \bibinfo {author} {\bibfnamefont {A.~O.}\ \bibnamefont {Sushkov}},\
		and\ \bibinfo {author} {\bibfnamefont {S.~K.}\ \bibnamefont {Lamoreaux}},\
	}\href {https://doi.org/10.1103/PhysRevLett.109.193003} {\bibfield  {journal}
		{\bibinfo  {journal} {Physical Review Letters}\ }\textbf {\bibinfo {volume}
			{109}},\ \bibinfo {pages} {193003} (\bibinfo {year} {2012})},\ \Eprint
	{https://arxiv.org/abs/1208.4420} {arXiv:1208.4420} \BibitemShut {NoStop}%
	\bibitem [{\citenamefont {Budker}\ \emph {et~al.}(2014)\citenamefont {Budker},
		\citenamefont {Graham}, \citenamefont {Ledbetter}, \citenamefont
		{Rajendran},\ and\ \citenamefont {Sushkov}}]{Budker2014}%
	\BibitemOpen
	\bibfield  {author} {\bibinfo {author} {\bibfnamefont {D.}~\bibnamefont
			{Budker}}, \bibinfo {author} {\bibfnamefont {P.~W.}\ \bibnamefont {Graham}},
		\bibinfo {author} {\bibfnamefont {M.}~\bibnamefont {Ledbetter}}, \bibinfo
		{author} {\bibfnamefont {S.}~\bibnamefont {Rajendran}},\ and\ \bibinfo
		{author} {\bibfnamefont {A.~O.}\ \bibnamefont {Sushkov}},\ }\href
	{https://doi.org/10.1103/PhysRevX.4.021030} {\bibfield  {journal} {\bibinfo
			{journal} {Physical Review X}\ }\textbf {\bibinfo {volume} {4}},\ \bibinfo
		{pages} {021030} (\bibinfo {year} {2014})}\BibitemShut {NoStop}%
	\bibitem [{\citenamefont {Aybas}\ \emph
		{et~al.}(2021{\natexlab{a}})\citenamefont {Aybas}, \citenamefont {Adam},
		\citenamefont {Blumenthal}, \citenamefont {Gramolin}, \citenamefont
		{Johnson}, \citenamefont {Kleyheeg}, \citenamefont {Afach}, \citenamefont
		{Blanchard}, \citenamefont {Centers}, \citenamefont {Garcon}, \citenamefont
		{Engler}, \citenamefont {Figueroa}, \citenamefont {Sendra}, \citenamefont
		{Wickenbrock}, \citenamefont {Lawson}, \citenamefont {Wang}, \citenamefont
		{Wu}, \citenamefont {Luo}, \citenamefont {Mani}, \citenamefont {Mauskopf},
		\citenamefont {Graham}, \citenamefont {Rajendran}, \citenamefont {Kimball},
		\citenamefont {Budker},\ and\ \citenamefont {Sushkov}}]{Aybas2021a}%
	\BibitemOpen
	\bibfield  {author} {\bibinfo {author} {\bibfnamefont {D.}~\bibnamefont
			{Aybas}}, \bibinfo {author} {\bibfnamefont {J.}~\bibnamefont {Adam}},
		\bibinfo {author} {\bibfnamefont {E.}~\bibnamefont {Blumenthal}}, \bibinfo
		{author} {\bibfnamefont {A.~V.}\ \bibnamefont {Gramolin}}, \bibinfo {author}
		{\bibfnamefont {D.}~\bibnamefont {Johnson}}, \bibinfo {author} {\bibfnamefont
			{A.}~\bibnamefont {Kleyheeg}}, \bibinfo {author} {\bibfnamefont
			{S.}~\bibnamefont {Afach}}, \bibinfo {author} {\bibfnamefont {J.~W.}\
			\bibnamefont {Blanchard}}, \bibinfo {author} {\bibfnamefont {G.~P.}\
			\bibnamefont {Centers}}, \bibinfo {author} {\bibfnamefont {A.}~\bibnamefont
			{Garcon}}, \bibinfo {author} {\bibfnamefont {M.}~\bibnamefont {Engler}},
		\bibinfo {author} {\bibfnamefont {N.~L.}\ \bibnamefont {Figueroa}}, \bibinfo
		{author} {\bibfnamefont {M.~G.}\ \bibnamefont {Sendra}}, \bibinfo {author}
		{\bibfnamefont {A.}~\bibnamefont {Wickenbrock}}, \bibinfo {author}
		{\bibfnamefont {M.}~\bibnamefont {Lawson}}, \bibinfo {author} {\bibfnamefont
			{T.}~\bibnamefont {Wang}}, \bibinfo {author} {\bibfnamefont {T.}~\bibnamefont
			{Wu}}, \bibinfo {author} {\bibfnamefont {H.}~\bibnamefont {Luo}}, \bibinfo
		{author} {\bibfnamefont {H.}~\bibnamefont {Mani}}, \bibinfo {author}
		{\bibfnamefont {P.}~\bibnamefont {Mauskopf}}, \bibinfo {author}
		{\bibfnamefont {P.~W.}\ \bibnamefont {Graham}}, \bibinfo {author}
		{\bibfnamefont {S.}~\bibnamefont {Rajendran}}, \bibinfo {author}
		{\bibfnamefont {D.~F.}\ \bibnamefont {Kimball}}, \bibinfo {author}
		{\bibfnamefont {D.}~\bibnamefont {Budker}},\ and\ \bibinfo {author}
		{\bibfnamefont {A.~O.}\ \bibnamefont {Sushkov}},\ }\href
	{https://doi.org/10.1103/PhysRevLett.126.141802} {\bibfield  {journal}
		{\bibinfo  {journal} {Physical Review Letters}\ }\textbf {\bibinfo {volume}
			{126}},\ \bibinfo {pages} {141802} (\bibinfo {year} {2021}{\natexlab{a}})},\
	\Eprint {https://arxiv.org/abs/2101.01241} {arXiv:2101.01241} \BibitemShut
	{NoStop}%
	\bibitem [{\citenamefont {Aybas}\ \emph
		{et~al.}(2021{\natexlab{b}})\citenamefont {Aybas}, \citenamefont {Bekker},
		\citenamefont {Blanchard}, \citenamefont {Budker}, \citenamefont {Centers},
		\citenamefont {Figueroa}, \citenamefont {Gramolin}, \citenamefont {{Jackson
				Kimball}}, \citenamefont {Wickenbrock},\ and\ \citenamefont
		{Sushkov}}]{Aybas2021b}%
	\BibitemOpen
	\bibfield  {author} {\bibinfo {author} {\bibfnamefont {D.}~\bibnamefont
			{Aybas}}, \bibinfo {author} {\bibfnamefont {H.}~\bibnamefont {Bekker}},
		\bibinfo {author} {\bibfnamefont {J.~W.}\ \bibnamefont {Blanchard}}, \bibinfo
		{author} {\bibfnamefont {D.}~\bibnamefont {Budker}}, \bibinfo {author}
		{\bibfnamefont {G.~P.}\ \bibnamefont {Centers}}, \bibinfo {author}
		{\bibfnamefont {N.~L.}\ \bibnamefont {Figueroa}}, \bibinfo {author}
		{\bibfnamefont {A.~V.}\ \bibnamefont {Gramolin}}, \bibinfo {author}
		{\bibfnamefont {D.~F.}\ \bibnamefont {{Jackson Kimball}}}, \bibinfo {author}
		{\bibfnamefont {A.}~\bibnamefont {Wickenbrock}},\ and\ \bibinfo {author}
		{\bibfnamefont {A.~O.}\ \bibnamefont {Sushkov}},\ }\href
	{https://doi.org/10.1088/2058-9565/abfbbc} {\bibfield  {journal} {\bibinfo
			{journal} {Quantum Science and Technology}\ }\textbf {\bibinfo {volume}
			{6}},\ \bibinfo {pages} {034007} (\bibinfo {year} {2021}{\natexlab{b}})},\
	\Eprint {https://arxiv.org/abs/2103.06284} {arXiv:2103.06284} \BibitemShut
	{NoStop}%
	\bibitem [{\citenamefont {Sushkov}\ \emph {et~al.}(2023)\citenamefont
		{Sushkov}, \citenamefont {Sushkov},\ and\ \citenamefont
		{Yaresko}}]{Sushkov2023}%
	\BibitemOpen
	\bibfield  {author} {\bibinfo {author} {\bibfnamefont {A.~O.}\ \bibnamefont
			{Sushkov}}, \bibinfo {author} {\bibfnamefont {O.~P.}\ \bibnamefont
			{Sushkov}},\ and\ \bibinfo {author} {\bibfnamefont {A.}~\bibnamefont
			{Yaresko}},\ }\href {https://arxiv.org/abs/2304.08461v1} {\bibfield
		{journal} {\bibinfo  {journal} {arXiv:2304.08461}\ } (\bibinfo {year}
		{2023})}\BibitemShut {NoStop}%
	\bibitem [{\citenamefont {Peccei}\ and\ \citenamefont
		{Quinn}(1977)}]{Peccei1977}%
	\BibitemOpen
	\bibfield  {author} {\bibinfo {author} {\bibfnamefont {R.~D.}\ \bibnamefont
			{Peccei}}\ and\ \bibinfo {author} {\bibfnamefont {H.~R.}\ \bibnamefont
			{Quinn}},\ }\href {https://doi.org/10.1103/PhysRevLett.38.1440} {\bibfield
		{journal} {\bibinfo  {journal} {Physical Review Letters}\ }\textbf {\bibinfo
			{volume} {38}},\ \bibinfo {pages} {1440} (\bibinfo {year}
		{1977})}\BibitemShut {NoStop}%
	\bibitem [{\citenamefont {Weinberg}(1978)}]{Weinberg1978}%
	\BibitemOpen
	\bibfield  {author} {\bibinfo {author} {\bibfnamefont {S.}~\bibnamefont
			{Weinberg}},\ }\href {https://doi.org/10.1103/PhysRevLett.40.223} {\bibfield
		{journal} {\bibinfo  {journal} {Physical Review Letters}\ }\textbf {\bibinfo
			{volume} {40}},\ \bibinfo {pages} {223} (\bibinfo {year} {1978})}\BibitemShut
	{NoStop}%
	\bibitem [{\citenamefont {Wilczek}(1978)}]{Wilczek1978}%
	\BibitemOpen
	\bibfield  {author} {\bibinfo {author} {\bibfnamefont {F.}~\bibnamefont
			{Wilczek}},\ }\href {https://doi.org/10.1103/PhysRevLett.40.279} {\bibfield
		{journal} {\bibinfo  {journal} {Physical Review Letters}\ }\textbf {\bibinfo
			{volume} {40}},\ \bibinfo {pages} {279} (\bibinfo {year} {1978})}\BibitemShut
	{NoStop}%
	\bibitem [{\citenamefont {Kim}\ and\ \citenamefont {Carosi}(2010)}]{Kim2010a}%
	\BibitemOpen
	\bibfield  {author} {\bibinfo {author} {\bibfnamefont {J.~E.}\ \bibnamefont
			{Kim}}\ and\ \bibinfo {author} {\bibfnamefont {G.}~\bibnamefont {Carosi}},\
	}\href {https://doi.org/10.1103/RevModPhys.82.557} {\bibfield  {journal}
		{\bibinfo  {journal} {Reviews of Modern Physics}\ }\textbf {\bibinfo {volume}
			{82}},\ \bibinfo {pages} {557} (\bibinfo {year} {2010})},\ \Eprint
	{https://arxiv.org/abs/0807.3125} {arXiv:0807.3125} \BibitemShut {NoStop}%
	\bibitem [{\citenamefont {Irastorza}\ and\ \citenamefont
		{Redondo}(2018)}]{Irastorza2018a}%
	\BibitemOpen
	\bibfield  {author} {\bibinfo {author} {\bibfnamefont {I.~G.}\ \bibnamefont
			{Irastorza}}\ and\ \bibinfo {author} {\bibfnamefont {J.}~\bibnamefont
			{Redondo}},\ }\href {https://doi.org/10.1016/j.ppnp.2018.05.003} {\bibfield
		{journal} {\bibinfo  {journal} {Progress in Particle and Nuclear Physics}\
		}\textbf {\bibinfo {volume} {102}},\ \bibinfo {pages} {89} (\bibinfo {year}
		{2018})},\ \Eprint {https://arxiv.org/abs/1801.08127} {arXiv:1801.08127}
	\BibitemShut {NoStop}%
	\bibitem [{\citenamefont {Sikivie}(1983)}]{Sikivie1983}%
	\BibitemOpen
	\bibfield  {author} {\bibinfo {author} {\bibfnamefont {P.}~\bibnamefont
			{Sikivie}},\ }\href {https://doi.org/10.1103/PhysRevLett.51.1415} {\bibfield
		{journal} {\bibinfo  {journal} {Physical Review Letters}\ }\textbf {\bibinfo
			{volume} {51}},\ \bibinfo {pages} {1415} (\bibinfo {year}
		{1983})}\BibitemShut {NoStop}%
	\bibitem [{\citenamefont {Braine}\ \emph {et~al.}(2020)\citenamefont {Braine},
		\citenamefont {Cervantes}, \citenamefont {Crisosto}, \citenamefont {Du},
		\citenamefont {Kimes}, \citenamefont {Rosenberg}, \citenamefont {Rybka},
		\citenamefont {Yang}, \citenamefont {Bowring}, \citenamefont {Chou},
		\citenamefont {Khatiwada}, \citenamefont {Sonnenschein}, \citenamefont
		{Wester}, \citenamefont {Carosi}, \citenamefont {Woollett}, \citenamefont
		{Duffy}, \citenamefont {Bradley}, \citenamefont {Boutan}, \citenamefont
		{Jones}, \citenamefont {Laroque}, \citenamefont {Oblath}, \citenamefont
		{Taubman}, \citenamefont {Clarke}, \citenamefont {Dove}, \citenamefont
		{Eddins}, \citenamefont {O'kelley}, \citenamefont {Nawaz}, \citenamefont
		{Siddiqi}, \citenamefont {Stevenson}, \citenamefont {Agrawal}, \citenamefont
		{Dixit}, \citenamefont {Gleason}, \citenamefont {Jois}, \citenamefont
		{Sikivie}, \citenamefont {Solomon}, \citenamefont {Sullivan}, \citenamefont
		{Tanner}, \citenamefont {Lentz}, \citenamefont {Daw}, \citenamefont
		{Buckley}, \citenamefont {Harrington}, \citenamefont {Henriksen},\ and\
		\citenamefont {Murch}}]{Braine2020}%
	\BibitemOpen
	\bibfield  {author} {\bibinfo {author} {\bibfnamefont {T.}~\bibnamefont
			{Braine}}, \bibinfo {author} {\bibfnamefont {R.}~\bibnamefont {Cervantes}},
		\bibinfo {author} {\bibfnamefont {N.}~\bibnamefont {Crisosto}}, \bibinfo
		{author} {\bibfnamefont {N.}~\bibnamefont {Du}}, \bibinfo {author}
		{\bibfnamefont {S.}~\bibnamefont {Kimes}}, \bibinfo {author} {\bibfnamefont
			{L.~J.}\ \bibnamefont {Rosenberg}}, \bibinfo {author} {\bibfnamefont
			{G.}~\bibnamefont {Rybka}}, \bibinfo {author} {\bibfnamefont
			{J.}~\bibnamefont {Yang}}, \bibinfo {author} {\bibfnamefont {D.}~\bibnamefont
			{Bowring}}, \bibinfo {author} {\bibfnamefont {A.~S.}\ \bibnamefont {Chou}},
		\bibinfo {author} {\bibfnamefont {R.}~\bibnamefont {Khatiwada}}, \bibinfo
		{author} {\bibfnamefont {A.}~\bibnamefont {Sonnenschein}}, \bibinfo {author}
		{\bibfnamefont {W.}~\bibnamefont {Wester}}, \bibinfo {author} {\bibfnamefont
			{G.}~\bibnamefont {Carosi}}, \bibinfo {author} {\bibfnamefont
			{N.}~\bibnamefont {Woollett}}, \bibinfo {author} {\bibfnamefont {L.~D.}\
			\bibnamefont {Duffy}}, \bibinfo {author} {\bibfnamefont {R.}~\bibnamefont
			{Bradley}}, \bibinfo {author} {\bibfnamefont {C.}~\bibnamefont {Boutan}},
		\bibinfo {author} {\bibfnamefont {M.}~\bibnamefont {Jones}}, \bibinfo
		{author} {\bibfnamefont {B.~H.}\ \bibnamefont {Laroque}}, \bibinfo {author}
		{\bibfnamefont {N.~S.}\ \bibnamefont {Oblath}}, \bibinfo {author}
		{\bibfnamefont {M.~S.}\ \bibnamefont {Taubman}}, \bibinfo {author}
		{\bibfnamefont {J.}~\bibnamefont {Clarke}}, \bibinfo {author} {\bibfnamefont
			{A.}~\bibnamefont {Dove}}, \bibinfo {author} {\bibfnamefont {A.}~\bibnamefont
			{Eddins}}, \bibinfo {author} {\bibfnamefont {S.~R.}\ \bibnamefont
			{O'kelley}}, \bibinfo {author} {\bibfnamefont {S.}~\bibnamefont {Nawaz}},
		\bibinfo {author} {\bibfnamefont {I.}~\bibnamefont {Siddiqi}}, \bibinfo
		{author} {\bibfnamefont {N.}~\bibnamefont {Stevenson}}, \bibinfo {author}
		{\bibfnamefont {A.}~\bibnamefont {Agrawal}}, \bibinfo {author} {\bibfnamefont
			{A.~V.}\ \bibnamefont {Dixit}}, \bibinfo {author} {\bibfnamefont {J.~R.}\
			\bibnamefont {Gleason}}, \bibinfo {author} {\bibfnamefont {S.}~\bibnamefont
			{Jois}}, \bibinfo {author} {\bibfnamefont {P.}~\bibnamefont {Sikivie}},
		\bibinfo {author} {\bibfnamefont {J.~A.}\ \bibnamefont {Solomon}}, \bibinfo
		{author} {\bibfnamefont {N.~S.}\ \bibnamefont {Sullivan}}, \bibinfo {author}
		{\bibfnamefont {D.~B.}\ \bibnamefont {Tanner}}, \bibinfo {author}
		{\bibfnamefont {E.}~\bibnamefont {Lentz}}, \bibinfo {author} {\bibfnamefont
			{E.~J.}\ \bibnamefont {Daw}}, \bibinfo {author} {\bibfnamefont {J.~H.}\
			\bibnamefont {Buckley}}, \bibinfo {author} {\bibfnamefont {P.~M.}\
			\bibnamefont {Harrington}}, \bibinfo {author} {\bibfnamefont {E.~A.}\
			\bibnamefont {Henriksen}},\ and\ \bibinfo {author} {\bibfnamefont {K.~W.}\
			\bibnamefont {Murch}},\ }\href
	{https://doi.org/10.1103/PhysRevLett.124.101303} {\bibfield  {journal}
		{\bibinfo  {journal} {Physical Review Letters}\ }\textbf {\bibinfo {volume}
			{124}},\ \bibinfo {pages} {101303} (\bibinfo {year} {2020})},\ \Eprint
	{https://arxiv.org/abs/1910.08638} {arXiv:1910.08638} \BibitemShut {NoStop}%
	\bibitem [{\citenamefont {Graham}\ and\ \citenamefont
		{Scherlis}(2018)}]{Graham2018a}%
	\BibitemOpen
	\bibfield  {author} {\bibinfo {author} {\bibfnamefont {P.~W.}\ \bibnamefont
			{Graham}}\ and\ \bibinfo {author} {\bibfnamefont {A.}~\bibnamefont
			{Scherlis}},\ }\href {https://doi.org/10.1103/PhysRevD.98.035017} {\bibfield
		{journal} {\bibinfo  {journal} {Physical Review D}\ }\textbf {\bibinfo
			{volume} {98}},\ \bibinfo {pages} {035017} (\bibinfo {year}
		{2018})}\BibitemShut {NoStop}%
	\bibitem [{\citenamefont {Ernst}\ \emph {et~al.}(2018)\citenamefont {Ernst},
		\citenamefont {Ringwald},\ and\ \citenamefont {Tamarit}}]{Ernst2018}%
	\BibitemOpen
	\bibfield  {author} {\bibinfo {author} {\bibfnamefont {A.}~\bibnamefont
			{Ernst}}, \bibinfo {author} {\bibfnamefont {A.}~\bibnamefont {Ringwald}},\
		and\ \bibinfo {author} {\bibfnamefont {C.}~\bibnamefont {Tamarit}},\ }\href
	{https://doi.org/10.1007/JHEP02(2018)103} {\bibfield  {journal} {\bibinfo
			{journal} {Journal of High Energy Physics}\ }\textbf {\bibinfo {volume}
			{2018}},\ \bibinfo {pages} {103} (\bibinfo {year} {2018})}\BibitemShut
	{NoStop}%
	\bibitem [{\citenamefont {Schutz}(2020)}]{Schutz2020}%
	\BibitemOpen
	\bibfield  {author} {\bibinfo {author} {\bibfnamefont {K.}~\bibnamefont
			{Schutz}},\ }\href {https://doi.org/10.1103/PhysRevD.101.123026} {\bibfield
		{journal} {\bibinfo  {journal} {Physical Review D}\ }\textbf {\bibinfo
			{volume} {101}},\ \bibinfo {pages} {123026} (\bibinfo {year} {2020})},\
	\Eprint {https://arxiv.org/abs/2001.05503} {arXiv:2001.05503} \BibitemShut
	{NoStop}%
	\bibitem [{\citenamefont {Graham}\ \emph {et~al.}(2015)\citenamefont {Graham},
		\citenamefont {Irastorza}, \citenamefont {Lamoreaux}, \citenamefont
		{Lindner},\ and\ \citenamefont {{Van Bibber}}}]{Graham2015a}%
	\BibitemOpen
	\bibfield  {author} {\bibinfo {author} {\bibfnamefont {P.~W.}\ \bibnamefont
			{Graham}}, \bibinfo {author} {\bibfnamefont {I.~G.}\ \bibnamefont
			{Irastorza}}, \bibinfo {author} {\bibfnamefont {S.~K.}\ \bibnamefont
			{Lamoreaux}}, \bibinfo {author} {\bibfnamefont {A.}~\bibnamefont {Lindner}},\
		and\ \bibinfo {author} {\bibfnamefont {K.~A.}\ \bibnamefont {{Van Bibber}}},\
	}\href {https://doi.org/10.1146/annurev-nucl-102014-022120} {\bibfield
		{journal} {\bibinfo  {journal} {Annual Review of Nuclear and Particle
				Science}\ }\textbf {\bibinfo {volume} {65}},\ \bibinfo {pages} {485}
		(\bibinfo {year} {2015})},\ \Eprint {https://arxiv.org/abs/1602.00039}
	{arXiv:1602.00039} \BibitemShut {NoStop}%
	\bibitem [{\citenamefont {Brubaker}\ \emph {et~al.}(2017)\citenamefont
		{Brubaker}, \citenamefont {Zhong}, \citenamefont {Gurevich}, \citenamefont
		{Cahn}, \citenamefont {Lamoreaux}, \citenamefont {Simanovskaia},
		\citenamefont {Root}, \citenamefont {Lewis}, \citenamefont {{Al Kenany}},
		\citenamefont {Backes}, \citenamefont {Urdinaran}, \citenamefont {Rapidis},
		\citenamefont {Shokair}, \citenamefont {{Van Bibber}}, \citenamefont
		{Palken}, \citenamefont {Malnou}, \citenamefont {Kindel}, \citenamefont
		{Anil}, \citenamefont {Lehnert},\ and\ \citenamefont
		{Carosi}}]{Brubaker2017b}%
	\BibitemOpen
	\bibfield  {author} {\bibinfo {author} {\bibfnamefont {B.~M.}\ \bibnamefont
			{Brubaker}}, \bibinfo {author} {\bibfnamefont {L.}~\bibnamefont {Zhong}},
		\bibinfo {author} {\bibfnamefont {Y.~V.}\ \bibnamefont {Gurevich}}, \bibinfo
		{author} {\bibfnamefont {S.~B.}\ \bibnamefont {Cahn}}, \bibinfo {author}
		{\bibfnamefont {S.~K.}\ \bibnamefont {Lamoreaux}}, \bibinfo {author}
		{\bibfnamefont {M.}~\bibnamefont {Simanovskaia}}, \bibinfo {author}
		{\bibfnamefont {J.~R.}\ \bibnamefont {Root}}, \bibinfo {author}
		{\bibfnamefont {S.~M.}\ \bibnamefont {Lewis}}, \bibinfo {author}
		{\bibfnamefont {S.}~\bibnamefont {{Al Kenany}}}, \bibinfo {author}
		{\bibfnamefont {K.~M.}\ \bibnamefont {Backes}}, \bibinfo {author}
		{\bibfnamefont {I.}~\bibnamefont {Urdinaran}}, \bibinfo {author}
		{\bibfnamefont {N.~M.}\ \bibnamefont {Rapidis}}, \bibinfo {author}
		{\bibfnamefont {T.~M.}\ \bibnamefont {Shokair}}, \bibinfo {author}
		{\bibfnamefont {K.~A.}\ \bibnamefont {{Van Bibber}}}, \bibinfo {author}
		{\bibfnamefont {D.~A.}\ \bibnamefont {Palken}}, \bibinfo {author}
		{\bibfnamefont {M.}~\bibnamefont {Malnou}}, \bibinfo {author} {\bibfnamefont
			{W.~F.}\ \bibnamefont {Kindel}}, \bibinfo {author} {\bibfnamefont {M.~A.}\
			\bibnamefont {Anil}}, \bibinfo {author} {\bibfnamefont {K.~W.}\ \bibnamefont
			{Lehnert}},\ and\ \bibinfo {author} {\bibfnamefont {G.}~\bibnamefont
			{Carosi}},\ }\href
	{https://doi.org/10.1103/PHYSREVLETT.118.061302/FIGURES/3/MEDIUM} {\bibfield
		{journal} {\bibinfo  {journal} {Physical Review Letters}\ }\textbf {\bibinfo
			{volume} {118}},\ \bibinfo {pages} {061302} (\bibinfo {year} {2017})},\
	\Eprint {https://arxiv.org/abs/1610.02580} {arXiv:1610.02580} \BibitemShut
	{NoStop}%
	\bibitem [{\citenamefont {McAllister}\ \emph {et~al.}(2017)\citenamefont
		{McAllister}, \citenamefont {Flower}, \citenamefont {Ivanov}, \citenamefont
		{Goryachev}, \citenamefont {Bourhill},\ and\ \citenamefont
		{Tobar}}]{McAllister2017}%
	\BibitemOpen
	\bibfield  {author} {\bibinfo {author} {\bibfnamefont {B.~T.}\ \bibnamefont
			{McAllister}}, \bibinfo {author} {\bibfnamefont {G.}~\bibnamefont {Flower}},
		\bibinfo {author} {\bibfnamefont {E.~N.}\ \bibnamefont {Ivanov}}, \bibinfo
		{author} {\bibfnamefont {M.}~\bibnamefont {Goryachev}}, \bibinfo {author}
		{\bibfnamefont {J.}~\bibnamefont {Bourhill}},\ and\ \bibinfo {author}
		{\bibfnamefont {M.~E.}\ \bibnamefont {Tobar}},\ }\href
	{https://doi.org/10.1016/j.dark.2017.09.010} {\bibfield  {journal} {\bibinfo
			{journal} {Physics of the Dark Universe}\ }\textbf {\bibinfo {volume} {18}},\
		\bibinfo {pages} {67} (\bibinfo {year} {2017})},\ \Eprint
	{https://arxiv.org/abs/1706.00209} {arXiv:1706.00209} \BibitemShut {NoStop}%
	\bibitem [{\citenamefont {Melc{\'{o}}n}\ \emph {et~al.}(2018)\citenamefont
		{Melc{\'{o}}n}, \citenamefont {Cuendis}, \citenamefont {Cogollos},
		\citenamefont {D{\'{i}}az-Morcillo}, \citenamefont {D{\"{o}}brich},
		\citenamefont {Gallego}, \citenamefont {Gimeno}, \citenamefont {Irastorza},
		\citenamefont {Lozano-Guerrero}, \citenamefont {Malbrunot}, \citenamefont
		{Navarro}, \citenamefont {Garay}, \citenamefont {Redondo}, \citenamefont
		{Vafeiadis},\ and\ \citenamefont {Wuensch}}]{Melcon2018}%
	\BibitemOpen
	\bibfield  {author} {\bibinfo {author} {\bibfnamefont {A.~{\'{A}}.}\
			\bibnamefont {Melc{\'{o}}n}}, \bibinfo {author} {\bibfnamefont {S.~A.}\
			\bibnamefont {Cuendis}}, \bibinfo {author} {\bibfnamefont {C.}~\bibnamefont
			{Cogollos}}, \bibinfo {author} {\bibfnamefont {A.}~\bibnamefont
			{D{\'{i}}az-Morcillo}}, \bibinfo {author} {\bibfnamefont {B.}~\bibnamefont
			{D{\"{o}}brich}}, \bibinfo {author} {\bibfnamefont {J.~D.}\ \bibnamefont
			{Gallego}}, \bibinfo {author} {\bibfnamefont {B.}~\bibnamefont {Gimeno}},
		\bibinfo {author} {\bibfnamefont {I.~G.}\ \bibnamefont {Irastorza}}, \bibinfo
		{author} {\bibfnamefont {A.~J.}\ \bibnamefont {Lozano-Guerrero}}, \bibinfo
		{author} {\bibfnamefont {C.}~\bibnamefont {Malbrunot}}, \bibinfo {author}
		{\bibfnamefont {P.}~\bibnamefont {Navarro}}, \bibinfo {author} {\bibfnamefont
			{C.~P.}\ \bibnamefont {Garay}}, \bibinfo {author} {\bibfnamefont
			{J.}~\bibnamefont {Redondo}}, \bibinfo {author} {\bibfnamefont
			{T.}~\bibnamefont {Vafeiadis}},\ and\ \bibinfo {author} {\bibfnamefont
			{W.}~\bibnamefont {Wuensch}},\ }\href
	{https://doi.org/10.1088/1475-7516/2018/05/040} {\bibfield  {journal}
		{\bibinfo  {journal} {Journal of Cosmology and Astroparticle Physics}\
		}\textbf {\bibinfo {volume} {2018}}\bibfield  {number} {\bibinfo  {number} {
				(5)},\ \bibinfo {pages} {040}},\ }\Eprint {https://arxiv.org/abs/1803.01243}
	{arXiv:1803.01243} \BibitemShut {NoStop}%
	\bibitem [{\citenamefont {Alesini}\ \emph {et~al.}(2019)\citenamefont
		{Alesini}, \citenamefont {Braggio}, \citenamefont {Carugno}, \citenamefont
		{Crescini}, \citenamefont {D'Agostino}, \citenamefont {{Di Gioacchino}},
		\citenamefont {{Di Vora}}, \citenamefont {Falferi}, \citenamefont {Gallo},
		\citenamefont {Gambardella}, \citenamefont {Gatti}, \citenamefont {Iannone},
		\citenamefont {Lamanna}, \citenamefont {Ligi}, \citenamefont {Lombardi},
		\citenamefont {Mezzena}, \citenamefont {Ortolan}, \citenamefont {Pengo},
		\citenamefont {Pompeo}, \citenamefont {Rettaroli}, \citenamefont {Ruoso},
		\citenamefont {Silva}, \citenamefont {Speake}, \citenamefont {Taffarello},\
		and\ \citenamefont {Tocci}}]{Alesini2019a}%
	\BibitemOpen
	\bibfield  {author} {\bibinfo {author} {\bibfnamefont {D.}~\bibnamefont
			{Alesini}}, \bibinfo {author} {\bibfnamefont {C.}~\bibnamefont {Braggio}},
		\bibinfo {author} {\bibfnamefont {G.}~\bibnamefont {Carugno}}, \bibinfo
		{author} {\bibfnamefont {N.}~\bibnamefont {Crescini}}, \bibinfo {author}
		{\bibfnamefont {D.}~\bibnamefont {D'Agostino}}, \bibinfo {author}
		{\bibfnamefont {D.}~\bibnamefont {{Di Gioacchino}}}, \bibinfo {author}
		{\bibfnamefont {R.}~\bibnamefont {{Di Vora}}}, \bibinfo {author}
		{\bibfnamefont {P.}~\bibnamefont {Falferi}}, \bibinfo {author} {\bibfnamefont
			{S.}~\bibnamefont {Gallo}}, \bibinfo {author} {\bibfnamefont
			{U.}~\bibnamefont {Gambardella}}, \bibinfo {author} {\bibfnamefont
			{C.}~\bibnamefont {Gatti}}, \bibinfo {author} {\bibfnamefont
			{G.}~\bibnamefont {Iannone}}, \bibinfo {author} {\bibfnamefont
			{G.}~\bibnamefont {Lamanna}}, \bibinfo {author} {\bibfnamefont
			{C.}~\bibnamefont {Ligi}}, \bibinfo {author} {\bibfnamefont {A.}~\bibnamefont
			{Lombardi}}, \bibinfo {author} {\bibfnamefont {R.}~\bibnamefont {Mezzena}},
		\bibinfo {author} {\bibfnamefont {A.}~\bibnamefont {Ortolan}}, \bibinfo
		{author} {\bibfnamefont {R.}~\bibnamefont {Pengo}}, \bibinfo {author}
		{\bibfnamefont {N.}~\bibnamefont {Pompeo}}, \bibinfo {author} {\bibfnamefont
			{A.}~\bibnamefont {Rettaroli}}, \bibinfo {author} {\bibfnamefont
			{G.}~\bibnamefont {Ruoso}}, \bibinfo {author} {\bibfnamefont
			{E.}~\bibnamefont {Silva}}, \bibinfo {author} {\bibfnamefont {C.~C.}\
			\bibnamefont {Speake}}, \bibinfo {author} {\bibfnamefont {L.}~\bibnamefont
			{Taffarello}},\ and\ \bibinfo {author} {\bibfnamefont {S.}~\bibnamefont
			{Tocci}},\ }\href {https://doi.org/10.1103/PhysRevD.99.101101} {\bibfield
		{journal} {\bibinfo  {journal} {Physical Review D}\ }\textbf {\bibinfo
			{volume} {99}},\ \bibinfo {pages} {101101} (\bibinfo {year} {2019})},\
	\Eprint {https://arxiv.org/abs/1903.06547} {arXiv:1903.06547} \BibitemShut
	{NoStop}%
	\bibitem [{\citenamefont {Lee}\ \emph {et~al.}(2020)\citenamefont {Lee},
		\citenamefont {Ahn}, \citenamefont {Choi}, \citenamefont {Ko},\ and\
		\citenamefont {Semertzidis}}]{Lee2020}%
	\BibitemOpen
	\bibfield  {author} {\bibinfo {author} {\bibfnamefont {S.}~\bibnamefont
			{Lee}}, \bibinfo {author} {\bibfnamefont {S.}~\bibnamefont {Ahn}}, \bibinfo
		{author} {\bibfnamefont {J.}~\bibnamefont {Choi}}, \bibinfo {author}
		{\bibfnamefont {B.~R.}\ \bibnamefont {Ko}},\ and\ \bibinfo {author}
		{\bibfnamefont {Y.~K.}\ \bibnamefont {Semertzidis}},\ }\href
	{https://doi.org/10.1103/PhysRevLett.124.101802} {\bibfield  {journal}
		{\bibinfo  {journal} {Physical Review Letters}\ }\textbf {\bibinfo {volume}
			{124}},\ \bibinfo {pages} {101802} (\bibinfo {year} {2020})},\ \Eprint
	{https://arxiv.org/abs/2001.05102} {arXiv:2001.05102} \BibitemShut {NoStop}%
	\bibitem [{\citenamefont {Sikivie}\ \emph {et~al.}(2013)\citenamefont
		{Sikivie}, \citenamefont {Sullivan},\ and\ \citenamefont
		{Tanner}}]{Sikivie2014b}%
	\BibitemOpen
	\bibfield  {author} {\bibinfo {author} {\bibfnamefont {P.}~\bibnamefont
			{Sikivie}}, \bibinfo {author} {\bibfnamefont {N.}~\bibnamefont {Sullivan}},\
		and\ \bibinfo {author} {\bibfnamefont {D.~B.}\ \bibnamefont {Tanner}},\
	}\href {https://doi.org/10.1103/PhysRevLett.112.131301} {\bibfield  {journal}
		{\bibinfo  {journal} {Physical Review Letters}\ }\textbf {\bibinfo {volume}
			{112}},\ \bibinfo {pages} {131301} (\bibinfo {year} {2013})}\BibitemShut
	{NoStop}%
	\bibitem [{\citenamefont {Chaudhuri}\ \emph {et~al.}(2015)\citenamefont
		{Chaudhuri}, \citenamefont {Graham}, \citenamefont {Irwin}, \citenamefont
		{Mardon}, \citenamefont {Rajendran},\ and\ \citenamefont
		{Zhao}}]{Chaudhuri2015}%
	\BibitemOpen
	\bibfield  {author} {\bibinfo {author} {\bibfnamefont {S.}~\bibnamefont
			{Chaudhuri}}, \bibinfo {author} {\bibfnamefont {P.~W.}\ \bibnamefont
			{Graham}}, \bibinfo {author} {\bibfnamefont {K.}~\bibnamefont {Irwin}},
		\bibinfo {author} {\bibfnamefont {J.}~\bibnamefont {Mardon}}, \bibinfo
		{author} {\bibfnamefont {S.}~\bibnamefont {Rajendran}},\ and\ \bibinfo
		{author} {\bibfnamefont {Y.}~\bibnamefont {Zhao}},\ }\href
	{https://doi.org/10.1103/PhysRevD.92.075012} {\bibfield  {journal} {\bibinfo
			{journal} {Physical Review D - Particles, Fields, Gravitation and Cosmology}\
		}\textbf {\bibinfo {volume} {92}},\ \bibinfo {pages} {075012} (\bibinfo
		{year} {2015})},\ \Eprint {https://arxiv.org/abs/1411.7382} {arXiv:1411.7382}
	\BibitemShut {NoStop}%
	\bibitem [{\citenamefont {Kahn}\ \emph {et~al.}(2016)\citenamefont {Kahn},
		\citenamefont {Safdi},\ and\ \citenamefont {Thaler}}]{Kahn2016}%
	\BibitemOpen
	\bibfield  {author} {\bibinfo {author} {\bibfnamefont {Y.}~\bibnamefont
			{Kahn}}, \bibinfo {author} {\bibfnamefont {B.~R.}\ \bibnamefont {Safdi}},\
		and\ \bibinfo {author} {\bibfnamefont {J.}~\bibnamefont {Thaler}},\ }\href
	{https://doi.org/10.1103/PhysRevLett.117.141801} {\bibfield  {journal}
		{\bibinfo  {journal} {Physical Review Letters}\ }\textbf {\bibinfo {volume}
			{117}},\ \bibinfo {pages} {141801} (\bibinfo {year} {2016})},\ \Eprint
	{https://arxiv.org/abs/1602.01086} {arXiv:1602.01086} \BibitemShut {NoStop}%
	\bibitem [{\citenamefont {Chaudhuri}\ \emph {et~al.}(2018)\citenamefont
		{Chaudhuri}, \citenamefont {Irwin}, \citenamefont {Graham},\ and\
		\citenamefont {Mardon}}]{Chaudhuri2018}%
	\BibitemOpen
	\bibfield  {author} {\bibinfo {author} {\bibfnamefont {S.}~\bibnamefont
			{Chaudhuri}}, \bibinfo {author} {\bibfnamefont {K.}~\bibnamefont {Irwin}},
		\bibinfo {author} {\bibfnamefont {P.~W.}\ \bibnamefont {Graham}},\ and\
		\bibinfo {author} {\bibfnamefont {J.}~\bibnamefont {Mardon}},\ }\href
	{http://arxiv.org/abs/1803.01627} {\bibfield  {journal} {\bibinfo  {journal}
			{arXiv:1803.01627}\ } (\bibinfo {year} {2018})},\ \Eprint
	{https://arxiv.org/abs/1803.01627} {arXiv:1803.01627} \BibitemShut {NoStop}%
	\bibitem [{\citenamefont {Ouellet}\ \emph {et~al.}(2019)\citenamefont
		{Ouellet}, \citenamefont {Salemi}, \citenamefont {Foster}, \citenamefont
		{Henning}, \citenamefont {Bogorad}, \citenamefont {Conrad}, \citenamefont
		{Formaggio}, \citenamefont {Kahn}, \citenamefont {Minervini}, \citenamefont
		{Radovinsky}, \citenamefont {Rodd}, \citenamefont {Safdi}, \citenamefont
		{Thaler}, \citenamefont {Winklehner},\ and\ \citenamefont
		{Winslow}}]{Ouellet2019}%
	\BibitemOpen
	\bibfield  {author} {\bibinfo {author} {\bibfnamefont {J.~L.}\ \bibnamefont
			{Ouellet}}, \bibinfo {author} {\bibfnamefont {C.~P.}\ \bibnamefont {Salemi}},
		\bibinfo {author} {\bibfnamefont {J.~W.}\ \bibnamefont {Foster}}, \bibinfo
		{author} {\bibfnamefont {R.}~\bibnamefont {Henning}}, \bibinfo {author}
		{\bibfnamefont {Z.}~\bibnamefont {Bogorad}}, \bibinfo {author} {\bibfnamefont
			{J.~M.}\ \bibnamefont {Conrad}}, \bibinfo {author} {\bibfnamefont {J.~A.}\
			\bibnamefont {Formaggio}}, \bibinfo {author} {\bibfnamefont {Y.}~\bibnamefont
			{Kahn}}, \bibinfo {author} {\bibfnamefont {J.}~\bibnamefont {Minervini}},
		\bibinfo {author} {\bibfnamefont {A.}~\bibnamefont {Radovinsky}}, \bibinfo
		{author} {\bibfnamefont {N.~L.}\ \bibnamefont {Rodd}}, \bibinfo {author}
		{\bibfnamefont {B.~R.}\ \bibnamefont {Safdi}}, \bibinfo {author}
		{\bibfnamefont {J.}~\bibnamefont {Thaler}}, \bibinfo {author} {\bibfnamefont
			{D.}~\bibnamefont {Winklehner}},\ and\ \bibinfo {author} {\bibfnamefont
			{L.}~\bibnamefont {Winslow}},\ }\href
	{https://doi.org/10.1103/PHYSREVLETT.122.121802/FIGURES/3/MEDIUM} {\bibfield
		{journal} {\bibinfo  {journal} {Physical Review Letters}\ }\textbf {\bibinfo
			{volume} {122}},\ \bibinfo {pages} {121802} (\bibinfo {year} {2019})},\
	\Eprint {https://arxiv.org/abs/1810.12257} {arXiv:1810.12257} \BibitemShut
	{NoStop}%
	\bibitem [{\citenamefont {Garcon}\ \emph {et~al.}(2019)\citenamefont {Garcon},
		\citenamefont {Blanchard}, \citenamefont {Centers}, \citenamefont {Figueroa},
		\citenamefont {Graham}, \citenamefont {{Jackson Kimball}}, \citenamefont
		{Rajendran}, \citenamefont {Sushkov}, \citenamefont {Stadnik}, \citenamefont
		{Wickenbrock}, \citenamefont {Wu},\ and\ \citenamefont
		{Budker}}]{Garcon2019b}%
	\BibitemOpen
	\bibfield  {author} {\bibinfo {author} {\bibfnamefont {A.}~\bibnamefont
			{Garcon}}, \bibinfo {author} {\bibfnamefont {J.~W.}\ \bibnamefont
			{Blanchard}}, \bibinfo {author} {\bibfnamefont {G.~P.}\ \bibnamefont
			{Centers}}, \bibinfo {author} {\bibfnamefont {N.~L.}\ \bibnamefont
			{Figueroa}}, \bibinfo {author} {\bibfnamefont {P.~W.}\ \bibnamefont
			{Graham}}, \bibinfo {author} {\bibfnamefont {D.~F.}\ \bibnamefont {{Jackson
					Kimball}}}, \bibinfo {author} {\bibfnamefont {S.}~\bibnamefont {Rajendran}},
		\bibinfo {author} {\bibfnamefont {A.~O.}\ \bibnamefont {Sushkov}}, \bibinfo
		{author} {\bibfnamefont {Y.~V.}\ \bibnamefont {Stadnik}}, \bibinfo {author}
		{\bibfnamefont {A.}~\bibnamefont {Wickenbrock}}, \bibinfo {author}
		{\bibfnamefont {T.}~\bibnamefont {Wu}},\ and\ \bibinfo {author}
		{\bibfnamefont {D.}~\bibnamefont {Budker}},\ }\href
	{https://doi.org/10.1126/sciadv.aax4539} {\bibfield  {journal} {\bibinfo
			{journal} {Science Advances}\ }\textbf {\bibinfo {volume} {5}},\ \bibinfo
		{pages} {eaax4539} (\bibinfo {year} {2019})},\ \Eprint
	{https://arxiv.org/abs/1902.04644} {arXiv:1902.04644} \BibitemShut {NoStop}%
	\bibitem [{\citenamefont {Wu}\ \emph {et~al.}(2019)\citenamefont {Wu},
		\citenamefont {Blanchard}, \citenamefont {Centers}, \citenamefont {Figueroa},
		\citenamefont {Garcon}, \citenamefont {Graham}, \citenamefont {Kimball},
		\citenamefont {Rajendran}, \citenamefont {Stadnik}, \citenamefont {Sushkov},
		\citenamefont {Wickenbrock},\ and\ \citenamefont {Budker}}]{Wu2019a}%
	\BibitemOpen
	\bibfield  {author} {\bibinfo {author} {\bibfnamefont {T.}~\bibnamefont
			{Wu}}, \bibinfo {author} {\bibfnamefont {J.~W.}\ \bibnamefont {Blanchard}},
		\bibinfo {author} {\bibfnamefont {G.~P.}\ \bibnamefont {Centers}}, \bibinfo
		{author} {\bibfnamefont {N.~L.}\ \bibnamefont {Figueroa}}, \bibinfo {author}
		{\bibfnamefont {A.}~\bibnamefont {Garcon}}, \bibinfo {author} {\bibfnamefont
			{P.~W.}\ \bibnamefont {Graham}}, \bibinfo {author} {\bibfnamefont {D.~F.}\
			\bibnamefont {Kimball}}, \bibinfo {author} {\bibfnamefont {S.}~\bibnamefont
			{Rajendran}}, \bibinfo {author} {\bibfnamefont {Y.~V.}\ \bibnamefont
			{Stadnik}}, \bibinfo {author} {\bibfnamefont {A.~O.}\ \bibnamefont
			{Sushkov}}, \bibinfo {author} {\bibfnamefont {A.}~\bibnamefont
			{Wickenbrock}},\ and\ \bibinfo {author} {\bibfnamefont {D.}~\bibnamefont
			{Budker}},\ }\href {https://doi.org/10.1103/PhysRevLett.122.191302}
	{\bibfield  {journal} {\bibinfo  {journal} {Physical Review Letters}\
		}\textbf {\bibinfo {volume} {122}},\ \bibinfo {pages} {191302} (\bibinfo
		{year} {2019})},\ \Eprint {https://arxiv.org/abs/1901.10843}
	{arXiv:1901.10843} \BibitemShut {NoStop}%
	\bibitem [{\citenamefont {Crescini}\ \emph {et~al.}(2020)\citenamefont
		{Crescini}, \citenamefont {Alesini}, \citenamefont {Braggio}, \citenamefont
		{Carugno}, \citenamefont {D'Agostino}, \citenamefont {{Di Gioacchino}},
		\citenamefont {Falferi}, \citenamefont {Gambardella}, \citenamefont {Gatti},
		\citenamefont {Iannone}, \citenamefont {Ligi}, \citenamefont {Lombardi},
		\citenamefont {Ortolan}, \citenamefont {Pengo}, \citenamefont {Ruoso},\ and\
		\citenamefont {Taffarello}}]{Crescini2020}%
	\BibitemOpen
	\bibfield  {author} {\bibinfo {author} {\bibfnamefont {N.}~\bibnamefont
			{Crescini}}, \bibinfo {author} {\bibfnamefont {D.}~\bibnamefont {Alesini}},
		\bibinfo {author} {\bibfnamefont {C.}~\bibnamefont {Braggio}}, \bibinfo
		{author} {\bibfnamefont {G.}~\bibnamefont {Carugno}}, \bibinfo {author}
		{\bibfnamefont {D.}~\bibnamefont {D'Agostino}}, \bibinfo {author}
		{\bibfnamefont {D.}~\bibnamefont {{Di Gioacchino}}}, \bibinfo {author}
		{\bibfnamefont {P.}~\bibnamefont {Falferi}}, \bibinfo {author} {\bibfnamefont
			{U.}~\bibnamefont {Gambardella}}, \bibinfo {author} {\bibfnamefont
			{C.}~\bibnamefont {Gatti}}, \bibinfo {author} {\bibfnamefont
			{G.}~\bibnamefont {Iannone}}, \bibinfo {author} {\bibfnamefont
			{C.}~\bibnamefont {Ligi}}, \bibinfo {author} {\bibfnamefont {A.}~\bibnamefont
			{Lombardi}}, \bibinfo {author} {\bibfnamefont {A.}~\bibnamefont {Ortolan}},
		\bibinfo {author} {\bibfnamefont {R.}~\bibnamefont {Pengo}}, \bibinfo
		{author} {\bibfnamefont {G.}~\bibnamefont {Ruoso}},\ and\ \bibinfo {author}
		{\bibfnamefont {L.}~\bibnamefont {Taffarello}},\ }\href
	{https://doi.org/10.1103/PhysRevLett.124.171801} {\bibfield  {journal}
		{\bibinfo  {journal} {Physical Review Letters}\ }\textbf {\bibinfo {volume}
			{124}},\ \bibinfo {pages} {171801} (\bibinfo {year} {2020})},\ \Eprint
	{https://arxiv.org/abs/2001.08940} {arXiv:2001.08940} \BibitemShut {NoStop}%
	\bibitem [{\citenamefont {Sikivie}(2021)}]{Sikivie2021}%
	\BibitemOpen
	\bibfield  {author} {\bibinfo {author} {\bibfnamefont {P.}~\bibnamefont
			{Sikivie}},\ }\href {https://doi.org/10.1103/RevModPhys.93.015004} {\bibfield
		{journal} {\bibinfo  {journal} {Reviews of Modern Physics}\ }\textbf
		{\bibinfo {volume} {93}},\ \bibinfo {pages} {015004} (\bibinfo {year}
		{2021})},\ \Eprint {https://arxiv.org/abs/2003.02206} {arXiv:2003.02206}
	\BibitemShut {NoStop}%
	\bibitem [{\citenamefont {Gramolin}\ \emph {et~al.}(2021)\citenamefont
		{Gramolin}, \citenamefont {Aybas}, \citenamefont {Johnson}, \citenamefont
		{Adam},\ and\ \citenamefont {Sushkov}}]{Gramolin2021}%
	\BibitemOpen
	\bibfield  {author} {\bibinfo {author} {\bibfnamefont {A.~V.}\ \bibnamefont
			{Gramolin}}, \bibinfo {author} {\bibfnamefont {D.}~\bibnamefont {Aybas}},
		\bibinfo {author} {\bibfnamefont {D.}~\bibnamefont {Johnson}}, \bibinfo
		{author} {\bibfnamefont {J.}~\bibnamefont {Adam}},\ and\ \bibinfo {author}
		{\bibfnamefont {A.~O.}\ \bibnamefont {Sushkov}},\ }\href
	{https://doi.org/10.1038/s41567-020-1006-6} {\bibfield  {journal} {\bibinfo
			{journal} {Nature Physics}\ }\textbf {\bibinfo {volume} {17}},\ \bibinfo
		{pages} {79} (\bibinfo {year} {2021})},\ \Eprint
	{https://arxiv.org/abs/2003.03348} {arXiv:2003.03348} \BibitemShut {NoStop}%
	\bibitem [{\citenamefont {{DMRadio Collaboration}}\ \emph
		{et~al.}(2023)\citenamefont {{DMRadio Collaboration}}, \citenamefont
		{AlShirawi}, \citenamefont {Bartram}, \citenamefont {Benabou}, \citenamefont
		{Brouwer}, \citenamefont {Chaudhuri}, \citenamefont {Cho}, \citenamefont
		{Corbin}, \citenamefont {Craddock}, \citenamefont {Droster}, \citenamefont
		{Foster}, \citenamefont {Fry}, \citenamefont {Graham}, \citenamefont
		{Henning}, \citenamefont {Irwin}, \citenamefont {Kadribasic}, \citenamefont
		{Kahn}, \citenamefont {Keller}, \citenamefont {Kolevatov}, \citenamefont
		{Kuenstner}, \citenamefont {Kurita}, \citenamefont {Leder}, \citenamefont
		{Li}, \citenamefont {Ouellet}, \citenamefont {Pappas}, \citenamefont
		{Phipps}, \citenamefont {Rapidis}, \citenamefont {Safdi}, \citenamefont
		{Salemi}, \citenamefont {Simanovskaia}, \citenamefont {Singh}, \citenamefont
		{van Assendelft}, \citenamefont {van Bibber}, \citenamefont {Wells},
		\citenamefont {Winslow}, \citenamefont {Wisniewski},\ and\ \citenamefont
		{Young}}]{DMRadioCollaboration2023}%
	\BibitemOpen
	\bibfield  {author} {\bibinfo {author} {\bibnamefont {{DMRadio
					Collaboration}}}, \bibinfo {author} {\bibfnamefont {A.}~\bibnamefont
			{AlShirawi}}, \bibinfo {author} {\bibfnamefont {C.}~\bibnamefont {Bartram}},
		\bibinfo {author} {\bibfnamefont {J.~N.}\ \bibnamefont {Benabou}}, \bibinfo
		{author} {\bibfnamefont {L.}~\bibnamefont {Brouwer}}, \bibinfo {author}
		{\bibfnamefont {S.}~\bibnamefont {Chaudhuri}}, \bibinfo {author}
		{\bibfnamefont {H.~M.}\ \bibnamefont {Cho}}, \bibinfo {author} {\bibfnamefont
			{J.}~\bibnamefont {Corbin}}, \bibinfo {author} {\bibfnamefont
			{W.}~\bibnamefont {Craddock}}, \bibinfo {author} {\bibfnamefont
			{A.}~\bibnamefont {Droster}}, \bibinfo {author} {\bibfnamefont {J.~W.}\
			\bibnamefont {Foster}}, \bibinfo {author} {\bibfnamefont {J.~T.}\
			\bibnamefont {Fry}}, \bibinfo {author} {\bibfnamefont {P.~W.}\ \bibnamefont
			{Graham}}, \bibinfo {author} {\bibfnamefont {R.}~\bibnamefont {Henning}},
		\bibinfo {author} {\bibfnamefont {K.~D.}\ \bibnamefont {Irwin}}, \bibinfo
		{author} {\bibfnamefont {F.}~\bibnamefont {Kadribasic}}, \bibinfo {author}
		{\bibfnamefont {Y.}~\bibnamefont {Kahn}}, \bibinfo {author} {\bibfnamefont
			{A.}~\bibnamefont {Keller}}, \bibinfo {author} {\bibfnamefont
			{R.}~\bibnamefont {Kolevatov}}, \bibinfo {author} {\bibfnamefont
			{S.}~\bibnamefont {Kuenstner}}, \bibinfo {author} {\bibfnamefont
			{N.}~\bibnamefont {Kurita}}, \bibinfo {author} {\bibfnamefont {A.~F.}\
			\bibnamefont {Leder}}, \bibinfo {author} {\bibfnamefont {D.}~\bibnamefont
			{Li}}, \bibinfo {author} {\bibfnamefont {J.~L.}\ \bibnamefont {Ouellet}},
		\bibinfo {author} {\bibfnamefont {K.~M.~W.}\ \bibnamefont {Pappas}}, \bibinfo
		{author} {\bibfnamefont {A.}~\bibnamefont {Phipps}}, \bibinfo {author}
		{\bibfnamefont {N.~M.}\ \bibnamefont {Rapidis}}, \bibinfo {author}
		{\bibfnamefont {B.~R.}\ \bibnamefont {Safdi}}, \bibinfo {author}
		{\bibfnamefont {C.~P.}\ \bibnamefont {Salemi}}, \bibinfo {author}
		{\bibfnamefont {M.}~\bibnamefont {Simanovskaia}}, \bibinfo {author}
		{\bibfnamefont {J.}~\bibnamefont {Singh}}, \bibinfo {author} {\bibfnamefont
			{E.~C.}\ \bibnamefont {van Assendelft}}, \bibinfo {author} {\bibfnamefont
			{K.}~\bibnamefont {van Bibber}}, \bibinfo {author} {\bibfnamefont
			{K.}~\bibnamefont {Wells}}, \bibinfo {author} {\bibfnamefont
			{L.}~\bibnamefont {Winslow}}, \bibinfo {author} {\bibfnamefont {W.~J.}\
			\bibnamefont {Wisniewski}},\ and\ \bibinfo {author} {\bibfnamefont {B.~A.}\
			\bibnamefont {Young}},\ }\href {https://arxiv.org/abs/2302.14084v1}
	{\bibfield  {journal} {\bibinfo  {journal} {arXiv:2302.14084}\ } (\bibinfo
		{year} {2023})}\BibitemShut {NoStop}%
	\bibitem [{\citenamefont {Graham}\ and\ \citenamefont
		{Rajendran}(2011)}]{Graham2011}%
	\BibitemOpen
	\bibfield  {author} {\bibinfo {author} {\bibfnamefont {P.~W.}\ \bibnamefont
			{Graham}}\ and\ \bibinfo {author} {\bibfnamefont {S.}~\bibnamefont
			{Rajendran}},\ }\href {https://doi.org/10.1103/PhysRevD.84.055013} {\bibfield
		{journal} {\bibinfo  {journal} {Physical Review D - Particles, Fields,
				Gravitation and Cosmology}\ }\textbf {\bibinfo {volume} {84}},\ \bibinfo
		{pages} {055013} (\bibinfo {year} {2011})},\ \Eprint
	{https://arxiv.org/abs/1101.2691} {arXiv:1101.2691} \BibitemShut {NoStop}%
	\bibitem [{\citenamefont {Ludlow}\ and\ \citenamefont
		{Sushkov}(2013)}]{Ludlow2013}%
	\BibitemOpen
	\bibfield  {author} {\bibinfo {author} {\bibfnamefont {J.~A.}\ \bibnamefont
			{Ludlow}}\ and\ \bibinfo {author} {\bibfnamefont {O.~P.}\ \bibnamefont
			{Sushkov}},\ }\href {https://doi.org/10.1088/0953-4075/46/8/085001}
	{\bibfield  {journal} {\bibinfo  {journal} {Journal of Physics B: Atomic,
				Molecular and Optical Physics}\ }\textbf {\bibinfo {volume} {46}},\ \bibinfo
		{pages} {085001} (\bibinfo {year} {2013})}\BibitemShut {NoStop}%
	\bibitem [{\citenamefont {Skripnikov}\ and\ \citenamefont
		{Titov}(2016)}]{Skripnikov2016}%
	\BibitemOpen
	\bibfield  {author} {\bibinfo {author} {\bibfnamefont {L.~V.}\ \bibnamefont
			{Skripnikov}}\ and\ \bibinfo {author} {\bibfnamefont {A.~V.}\ \bibnamefont
			{Titov}},\ }\href {https://doi.org/10.1063/1.4959973} {\bibfield  {journal}
		{\bibinfo  {journal} {Journal of Chemical Physics}\ }\textbf {\bibinfo
			{volume} {145}},\ \bibinfo {pages} {054115} (\bibinfo {year}
		{2016})}\BibitemShut {NoStop}%
	\bibitem [{\citenamefont {Flambaum}\ and\ \citenamefont
		{Samsonov}(2020)}]{Flambaum2020}%
	\BibitemOpen
	\bibfield  {author} {\bibinfo {author} {\bibfnamefont {V.~V.}\ \bibnamefont
			{Flambaum}}\ and\ \bibinfo {author} {\bibfnamefont {I.~B.}\ \bibnamefont
			{Samsonov}},\ }\href {https://doi.org/10.1103/PhysRevResearch.2.023042}
	{\bibfield  {journal} {\bibinfo  {journal} {Physical Review Research}\
		}\textbf {\bibinfo {volume} {2}},\ \bibinfo {pages} {023042} (\bibinfo {year}
		{2020})}\BibitemShut {NoStop}%
	\bibitem [{\citenamefont {Dalton}\ \emph {et~al.}(2023)\citenamefont {Dalton},
		\citenamefont {Flambaum},\ and\ \citenamefont {Mansour}}]{Dalton2023}%
	\BibitemOpen
	\bibfield  {author} {\bibinfo {author} {\bibfnamefont {F.}~\bibnamefont
			{Dalton}}, \bibinfo {author} {\bibfnamefont {V.~V.}\ \bibnamefont
			{Flambaum}},\ and\ \bibinfo {author} {\bibfnamefont {A.~J.}\ \bibnamefont
			{Mansour}},\ }\href {https://doi.org/10.1103/PhysRevC.107.035502} {\bibfield
		{journal} {\bibinfo  {journal} {Physical Review C}\ }\textbf {\bibinfo
			{volume} {107}},\ \bibinfo {pages} {035502} (\bibinfo {year} {2023})},\
	\Eprint {https://arxiv.org/abs/2302.00214} {arXiv:2302.00214} \BibitemShut
	{NoStop}%
	\bibitem [{\citenamefont {Ahlefeldt}\ \emph {et~al.}(2016)\citenamefont
		{Ahlefeldt}, \citenamefont {Hush},\ and\ \citenamefont
		{Sellars}}]{Ahlefeldt2016}%
	\BibitemOpen
	\bibfield  {author} {\bibinfo {author} {\bibfnamefont {R.~L.}\ \bibnamefont
			{Ahlefeldt}}, \bibinfo {author} {\bibfnamefont {M.~R.}\ \bibnamefont
			{Hush}},\ and\ \bibinfo {author} {\bibfnamefont {M.~J.}\ \bibnamefont
			{Sellars}},\ }\href {https://doi.org/10.1103/PhysRevLett.117.250504}
	{\bibfield  {journal} {\bibinfo  {journal} {Physical Review Letters}\
		}\textbf {\bibinfo {volume} {117}},\ \bibinfo {pages} {250504} (\bibinfo
		{year} {2016})},\ \Eprint {https://arxiv.org/abs/1601.05013}
	{arXiv:1601.05013} \BibitemShut {NoStop}%
	\bibitem [{\citenamefont {Ahlefeldt}\ \emph
		{et~al.}(2013{\natexlab{a}})\citenamefont {Ahlefeldt}, \citenamefont {Zhong},
		\citenamefont {Bartholomew},\ and\ \citenamefont {Sellars}}]{Ahlefeldt2013b}%
	\BibitemOpen
	\bibfield  {author} {\bibinfo {author} {\bibfnamefont {R.~L.}\ \bibnamefont
			{Ahlefeldt}}, \bibinfo {author} {\bibfnamefont {M.}~\bibnamefont {Zhong}},
		\bibinfo {author} {\bibfnamefont {J.~G.}\ \bibnamefont {Bartholomew}},\ and\
		\bibinfo {author} {\bibfnamefont {M.~J.}\ \bibnamefont {Sellars}},\ }\href
	{https://doi.org/10.1016/j.jlumin.2013.04.046} {\bibfield  {journal}
		{\bibinfo  {journal} {Journal of Luminescence}\ }\textbf {\bibinfo {volume}
			{143}},\ \bibinfo {pages} {193} (\bibinfo {year}
		{2013}{\natexlab{a}})}\BibitemShut {NoStop}%
	\bibitem [{\citenamefont {Tambornino}\ \emph {et~al.}(2014)\citenamefont
		{Tambornino}, \citenamefont {Bielec},\ and\ \citenamefont
		{Hoch}}]{Tambornino2014}%
	\BibitemOpen
	\bibfield  {author} {\bibinfo {author} {\bibfnamefont {F.}~\bibnamefont
			{Tambornino}}, \bibinfo {author} {\bibfnamefont {P.}~\bibnamefont {Bielec}},\
		and\ \bibinfo {author} {\bibfnamefont {C.}~\bibnamefont {Hoch}},\ }\href
	{https://doi.org/10.1107/S1600536814010307} {\bibfield  {journal} {\bibinfo
			{journal} {Acta Crystallographica Section E: Structure Reports Online}\
		}\textbf {\bibinfo {volume} {70}},\ \bibinfo {pages} {i27} (\bibinfo {year}
		{2014})}\BibitemShut {NoStop}%
	\bibitem [{\citenamefont {Smith}\ \emph {et~al.}(2022)\citenamefont {Smith},
		\citenamefont {Reid}, \citenamefont {Sellars},\ and\ \citenamefont
		{Ahlefeldt}}]{Smith2022}%
	\BibitemOpen
	\bibfield  {author} {\bibinfo {author} {\bibfnamefont {K.~M.}\ \bibnamefont
			{Smith}}, \bibinfo {author} {\bibfnamefont {M.~F.}\ \bibnamefont {Reid}},
		\bibinfo {author} {\bibfnamefont {M.~J.}\ \bibnamefont {Sellars}},\ and\
		\bibinfo {author} {\bibfnamefont {R.~L.}\ \bibnamefont {Ahlefeldt}},\ }\href
	{https://doi.org/10.1103/PhysRevB.105.125141} {\bibfield  {journal} {\bibinfo
			{journal} {Physical Review B}\ }\textbf {\bibinfo {volume} {105}},\ \bibinfo
		{pages} {125141} (\bibinfo {year} {2022})},\ \Eprint
	{https://arxiv.org/abs/2110.03896} {arXiv:2110.03896} \BibitemShut {NoStop}%
	\bibitem [{\citenamefont {Ahlefeldt}\ \emph
		{et~al.}(2013{\natexlab{b}})\citenamefont {Ahlefeldt}, \citenamefont
		{Manson},\ and\ \citenamefont {Sellars}}]{Ahlefeldt2013c}%
	\BibitemOpen
	\bibfield  {author} {\bibinfo {author} {\bibfnamefont {R.~L.}\ \bibnamefont
			{Ahlefeldt}}, \bibinfo {author} {\bibfnamefont {N.~B.}\ \bibnamefont
			{Manson}},\ and\ \bibinfo {author} {\bibfnamefont {M.~J.}\ \bibnamefont
			{Sellars}},\ }\href {https://doi.org/10.1016/j.jlumin.2011.12.036} {\bibfield
		{journal} {\bibinfo  {journal} {Journal of Luminescence}\ }\textbf {\bibinfo
			{volume} {133}},\ \bibinfo {pages} {152} (\bibinfo {year}
		{2013}{\natexlab{b}})}\BibitemShut {NoStop}%
	\bibitem [{\citenamefont {Graham}\ and\ \citenamefont
		{Rajendran}(2013)}]{Graham2013}%
	\BibitemOpen
	\bibfield  {author} {\bibinfo {author} {\bibfnamefont {P.~W.}\ \bibnamefont
			{Graham}}\ and\ \bibinfo {author} {\bibfnamefont {S.}~\bibnamefont
			{Rajendran}},\ }\href {https://doi.org/10.1103/PhysRevD.88.035023} {\bibfield
		{journal} {\bibinfo  {journal} {Physical Review D - Particles, Fields,
				Gravitation and Cosmology}\ }\textbf {\bibinfo {volume} {88}},\ \bibinfo
		{pages} {035023} (\bibinfo {year} {2013})},\ \Eprint
	{https://arxiv.org/abs/1306.6088} {arXiv:1306.6088} \BibitemShut {NoStop}%
	\bibitem [{\citenamefont {Sushkov}\ \emph {et~al.}(2010)\citenamefont
		{Sushkov}, \citenamefont {Eckel},\ and\ \citenamefont
		{Lamoreaux}}]{Sushkov2010}%
	\BibitemOpen
	\bibfield  {author} {\bibinfo {author} {\bibfnamefont {A.~O.}\ \bibnamefont
			{Sushkov}}, \bibinfo {author} {\bibfnamefont {S.}~\bibnamefont {Eckel}},\
		and\ \bibinfo {author} {\bibfnamefont {S.~K.}\ \bibnamefont {Lamoreaux}},\
	}\href {https://doi.org/10.1103/PhysRevA.81.022104} {\bibfield  {journal}
		{\bibinfo  {journal} {Physical Review A}\ }\textbf {\bibinfo {volume} {81}},\
		\bibinfo {pages} {022104} (\bibinfo {year} {2010})}\BibitemShut {NoStop}%
\end{thebibliography}
%\end{document}


\end{document}